\documentclass[%
reprint,
groupedaddress,
 amsmath,amssymb,
 aps,
pra,
]{revtex4-1}
\usepackage{graphicx}
\usepackage{upgreek}

\begin{document}

\preprint{APS/123-QED}

\title{Quantum Simulation of Ultrafast Dynamics Using Trapped Ultracold Atoms}

\author{Ruwan Senaratne}
\thanks{These authors contributed equally.}
\author{Shankari V. Rajagopal}
\thanks{These authors contributed equally.}
\author{Toshihiko Shimasaki}
\author{Peter E. Dotti}
\author{Kurt M. Fujiwara}
\author{Kevin Singh}
\author{Zachary A. Geiger}
\author{David M.~Weld}
\email{Correspondence and requests for materials should be addressed to D.M.W. (email: weld@ucsb.edu).}
\affiliation{University of California and California Institute for Quantum Emulation, Santa Barbara CA 93106}

\begin{abstract}
Ultrafast electronic dynamics are typically studied using pulsed lasers. We demonstrate a complementary experimental approach: quantum simulation of ultrafast dynamics using trapped ultracold atoms. Counter-intuitively, this technique emulates some of the fastest processes in atomic physics with some of the slowest, leading to a temporal magnification factor of up to twelve orders of magnitude. In these experiments, time-varying forces on neutral atoms in the ground state of a tunable optical trap emulate the electric fields of a pulsed laser acting on bound charged particles. We demonstrate the correspondence with ultrafast science by a sequence of experiments: nonlinear spectroscopy of a many-body bound state, control of the excitation spectrum by potential shaping, observation of sub-cycle unbinding dynamics during strong few-cycle pulses, and direct measurement of carrier-envelope phase dependence of the response to an ultrafast-equivalent pulse. These results establish cold atom quantum simulation as a complementary tool for studying ultrafast dynamics. 
\end{abstract}

\maketitle

The study of ultrafast-equivalent electronic and vibrational dynamics is a natural but largely unexplored application of cold-atom quantum simulation techniques~\cite{lewenstein-wavepacketdynamics,saenz-attosecondsimulatorPRA,holthaus-strongfieldsim,sengstock-moleculesimulator,adppaper}. Quantum simulation experiments often rely on an analogy between trapped neutral atoms and electrons in matter~\cite{lewenstein-review, blochdalibardQSreview,bloch2017review}. Although these two systems have vastly different energy densities and constituents which differ in mass and charge, they can often be described by equivalent Hamiltonians, which give rise to equivalent physics. This analogy has been used to explore equilibrium solid-state phenomena from Mott insulators to antiferromagnets~\cite{greinerSFMI,HuletAF}, and dynamical phenomena from Bloch oscillations to many-body localization~\cite{salomon-blochoscs,bloch-mblscience}. Here we extend this analogy to quantum simulation of ultrafast dynamics, with the aim of realizing an alternate experimental approach to open questions in a vibrant and expanding area of science~\cite{krausz-attosecond-review,corkumkrauszreview,clusters}, testing approximate theories~\cite{Keldysh1964,Faisal1973,Reiss1980,corkum-model,lewenstein-hhg}, and pushing into experimentally unexplored  regimes. 

\begin{figure}[t]
\centering
\includegraphics[width=\columnwidth]{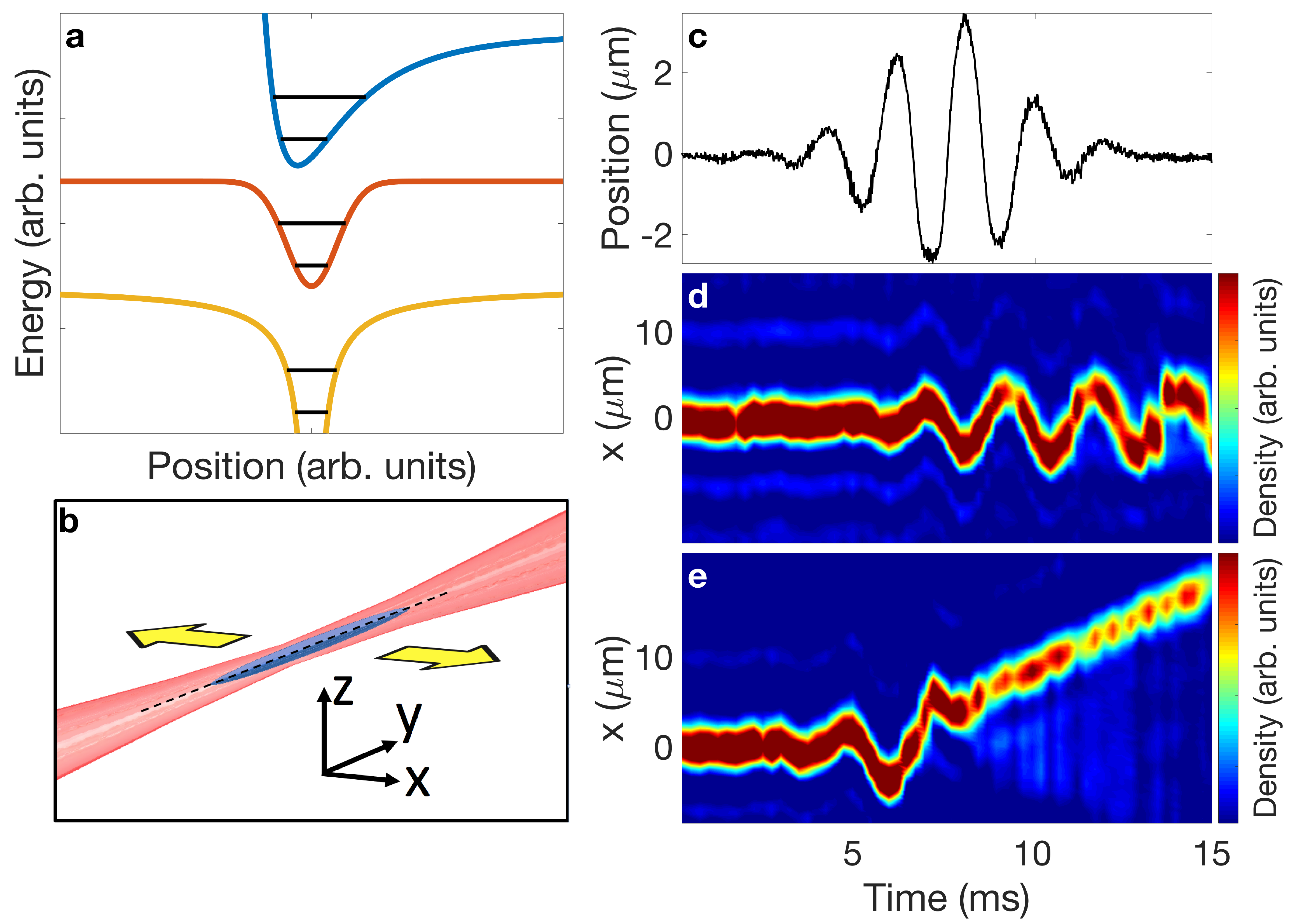}
\caption{\textbf{Quantum simulation of ultrafast dynamics.} \textbf{a:}~Schematic bound states of Lennard-Jones, Gaussian, and $1/r$ potentials (offset for clarity). \textbf{b:}~Diagram of optical trap (red), which is shaken in the $\mathbf{\hat{x}}$ direction to generate inertial forces on the condensate (blue). \textbf{c:}~Measured trap position $\alpha(t)$ during a pulse.
\textbf{d:}~Response to a weak pulse. Colourmap shows density distribution after time-of-flight as a function of time. Pulse carrier frequency is 450 Hz, pulse envelope width is 3.76 ms, pulse amplitude is 0.6 $\upmu$m, and carrier-envelope phase is 0, as defined in Eq.~\ref{eqn:pulse}.
\textbf{e:}~Response to a stronger pulse. Unbinding occurs near 8 ms, after which the atoms propagate with constant velocity. Pulse amplitude is 2.4 $\upmu$m. All other pulse parameters are identical to those in \textbf{d}.}
\label{fig:summary}
\end{figure}
The quantum simulator we describe consists of an artificial atom or molecule made from a trapped quantum gas.  The analogue of the atomic or molecular binding potential is the tunable AC Stark potential of an optical trap, and the analogue of the pulsed laser's electric field is an inertial force arising from rapid trap translation. The time-dependent Gross-Pitaevskii equation describing the evolution of the condensate wavefunction $\mathit{\Psi}(\boldsymbol{\mathrm{r}},t)$ is~\cite{lewenstein-wavepacketdynamics}
\begin{equation}
\left[-\mathrm{i}\hbar\partial_t -\frac{\hbar^2\nabla^2}{2m}+V(\boldsymbol{\mathrm{r}}+\alpha(t)\boldsymbol{\mathrm{\hat{x}}})+gN|\mathit{\Psi|}^2\right]\mathit{\Psi}=0,
\label{GPE}
\end{equation}
where $m$ is the atomic mass,  $g=4\mathrm{\pi}\hbar^2a_{\mathrm{s}}/m$ parameterizes interactions among $N$ atoms, $a_{\mathrm{s}}$ is the scattering length, and the optical potential $V(\boldsymbol{\mathrm{r}})$ is shaken in the $\boldsymbol{\mathrm{\hat{x}}}$ direction with waveform $\alpha(t)$. Crucially, the same equation also describes the evolution of the wavefunction of an atomic electron interacting with a linearly-polarized laser field in the Kramers-Henneberger frame of reference~\cite{lewenstein-wavepacketdynamics}, taking $m$ to be the electron mass, $V$ the nuclear potential including screening effects, and $g\rightarrow 0$. For dipolar excitations like those in the experiments we present, the impact of the atoms' nonzero $g$ is minimized due to Kohn's theorem. Very similar dynamics have been theoretically predicted~\cite{lewenstein-wavepacketdynamics} for the cold-atom and ultrafast realizations of Eq.~\ref{GPE}. A closely related equivalence is described in \cite{saenz-attosecondsimulatorPRA}. This equivalence between the evolution of condensate and electron wavefunctions motivates cold atom quantum simulation of ultrafast dynamics, much as the Bose-Hubbard model motivated early quantum simulation of Mott insulators~\cite{greinerSFMI}. 

Though very little of the growing body of quantum simulation work has addressed ultrafast phenomena, a robust toolkit exists for controlling and measuring excitations in trapped gases. Collective excitations in Bose condensates were a major focus of early experimental and theoretical research~\cite{jin-collectiveexc,ketterle-collectiveexcitations,inguscioexcitations,stringari-orig,shlyapnikov-excitations,graham-anisotropictrapexc}, and the analogy between degenerate trapped gases and individual atoms was noted at that time~\cite{esry97,walsworth97,lewenstein-wavepacketdynamics}.  Ultrafast probes have recently been used to study many-body dynamics in Rydberg atoms~\cite{ultrafastrydberg}, and recent theoretical proposals have suggested the use of cold atoms to simulate ultrafast dynamics in atoms~\cite{saenz-attosecondsimulatorPRA,adppaper}, molecules~\cite{sengstock-moleculesimulator}, and solids~\cite{holthaus-strongfieldsim}. 

Cold gases offer unique capabilities for dynamical quantum simulation. Due to the extremely low energy scales, the dynamics are slowed, or magnified, with respect to atomic or molecular timescales by as much as twelve orders of magnitude, allowing the observation of ultrafast-equivalent processes in ultra-slow-motion~\cite{saenz-attosecondsimulatorPRA}.  This extreme temporal magnification --- quantum gas chronoscopy --- enables simple and complete control over all parameters of an applied force pulse, as well as straightforward measurement of the artificial atom's or molecule's response, with time resolution much faster than all relevant dynamics.  The excitation spectrum itself can also be controlled by trap shaping. Using this toolkit of capabilities, we demonstrate experimentally that cold atom quantum simulation can be used to probe complex phenomena of ultrafast science such as the effect of carrier-envelope phase and pulse intensity on unbinding dynamics. These experiments demonstrate a new application for cold-atom quantum simulation and establish a potentially fruitful connection between ultrafast and ultracold atomic physics.

\section{Results}

\textbf{Ultrafast-equivalent pulse synthesis.} The experiments we describe use a Bose condensate of
$N\!\simeq\! 20,000$ atoms of $^{84}$Sr~\cite{strontium84BEC}, with a scattering length $a_{\mathrm{s}}\!\simeq\!6.5~\mathrm{nm}$. Rapid trap translation gives rise to time-dependent inertial forces designed to have the same approximate functional form, and the same effect of driving dipole-allowed transitions, as the electric field of an ultrafast pulsed laser. This is achieved by applying a trap which depends on $x$ and $t$ as \mbox{$V(x,t) = -V_\text{trap} \times \exp\left[-2 (x-\alpha(t))^2/w^2\right]$}, where $w$ is the 1/$\mathrm{e}^2$ trap waist and 
\begin{equation}
\alpha(t) = A\  {\rm sech} \left[ \eta \left( t - t_0 \right) \right]  \sin \left[ 2 \pi f \left( t - t_0 \right) + \phi  + \mathrm{\pi} \right].
\label{eqn:pulse}
\end{equation}
Control over the pulse is effectively arbitrary; variable parameters include amplitude $A$, carrier frequency $f$, pulse full-width at half-maximum $\tau=(2 \ln{(2+\sqrt{3})})/\eta$, and carrier-envelope phase $\phi$.  The measured trap centre translation as a function of time during a typical pulse is shown in Fig.~\ref{fig:summary}c. All data reported here use pulse amplitudes well below the trap width.  The effective Keldysh parameter in such an experiment is $\gamma_{\mathrm{K}} = \sqrt{V_\text{trap}/2U_{\mathrm{p}}}$, where the optical trap depth $V_\text{trap}$ corresponds to the ionization energy and the ponderomotive potential $U_{\mathrm{p}} \simeq  m \overline{\dot{\alpha}^2}/2$ is the time-averaged kinetic energy imparted to the atoms by the pulse. The use of inertial forces enables realization of Keldysh parameters of order unity and greater. Keldysh parameters much less than 1 could be straightforwardly attained by using a time-varying optical potential gradient rather than trap motion to apply the simulated electric field.

\begin{figure}[t!]
\centering
\includegraphics[width=0.95\linewidth]{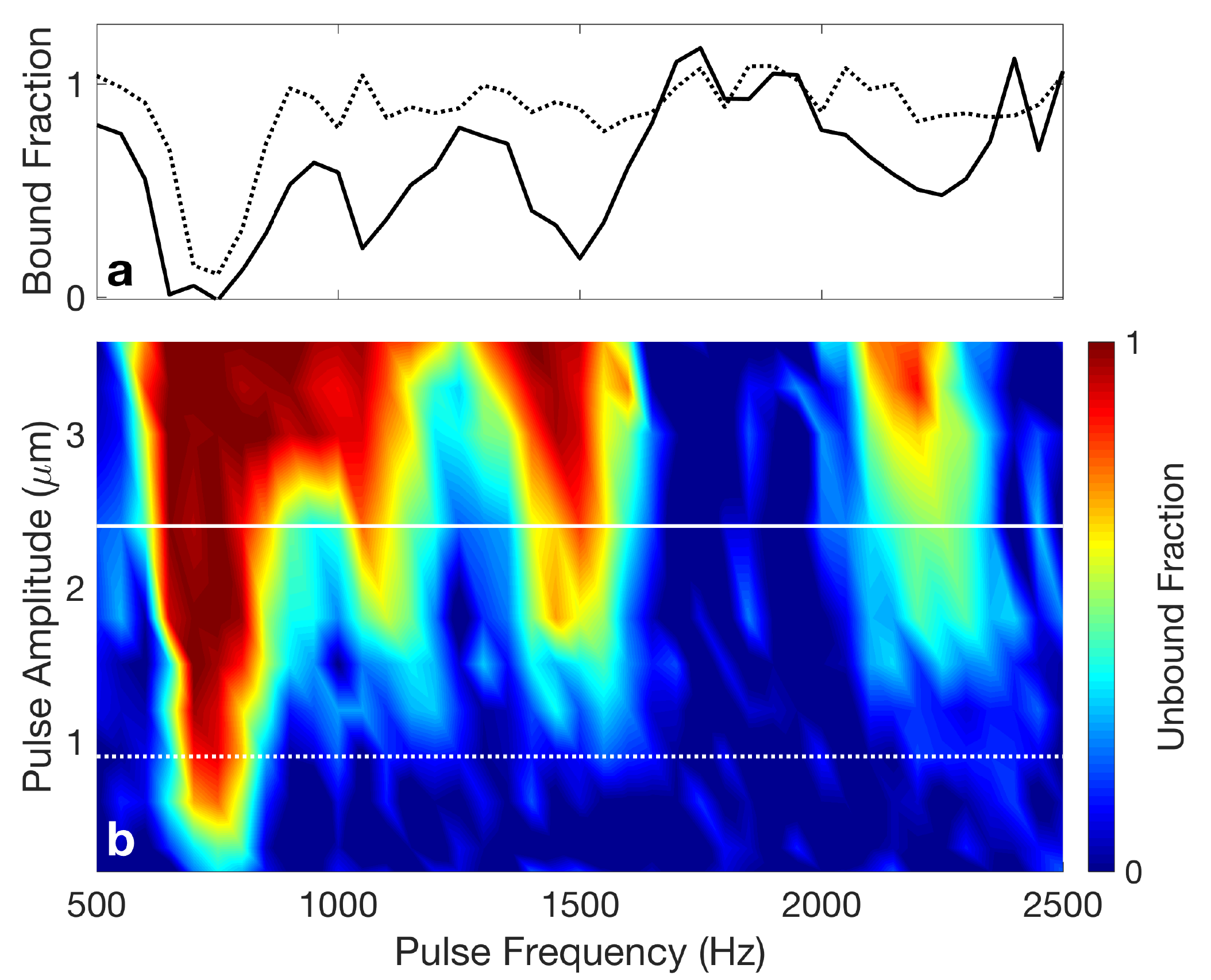}
\caption{\textbf{Spectroscopy of the quantum simulator.} \textbf{a:}~Remaining bound fraction as a function of carrier frequency for pulses with $\tau =250$~ms, $\phi=\mathrm{\pi}$ and amplitudes of 0.9~$\upmu$m (dotted) and 2.4~$\upmu$m (solid). Note the emergence of higher-order peaks and power broadening at larger amplitudes. \textbf{b:}~Unbound fraction as a function of applied pulse frequency and amplitude, for a 250~ms pulse. Lines indicate cuts plotted in panel \textbf{a}. }
\label{fig:spec}
\end{figure}

\textbf{Spectroscopy of tunable collective excitations.} We performed initial spectroscopic characterization of our quantum simulator by applying pulses of constant length much greater than a drive period and variable carrier frequency $f$.  After each pulse, the atoms that had not been unbound from the trap were counted with absorption imaging.  The resulting plots of bound fraction versus pulse frequency characterize the collective excitation spectra of the trapped condensate. Nonlinear effects are straightforwardly probed by increasing the pulse intensity.

Excitation spectra for one particular trap are shown in the top panel of Fig. \ref{fig:spec}. The resonance at $\sim$750 Hz corresponds to dipole oscillation in the trap and is at the same frequency as the resonance for a non-interacting gas.  As the pulse amplitude is increased, higher modes are excited and power broadening is observed. Since our trap is deeply in the Thomas-Fermi regime, these modes are anharmonic and strongly collective. The bottom panel of Fig.~\ref{fig:spec} shows a 2D amplitude-frequency spectrogram.

\begin{figure}[t!]
\centering
\includegraphics[width=0.95\linewidth]{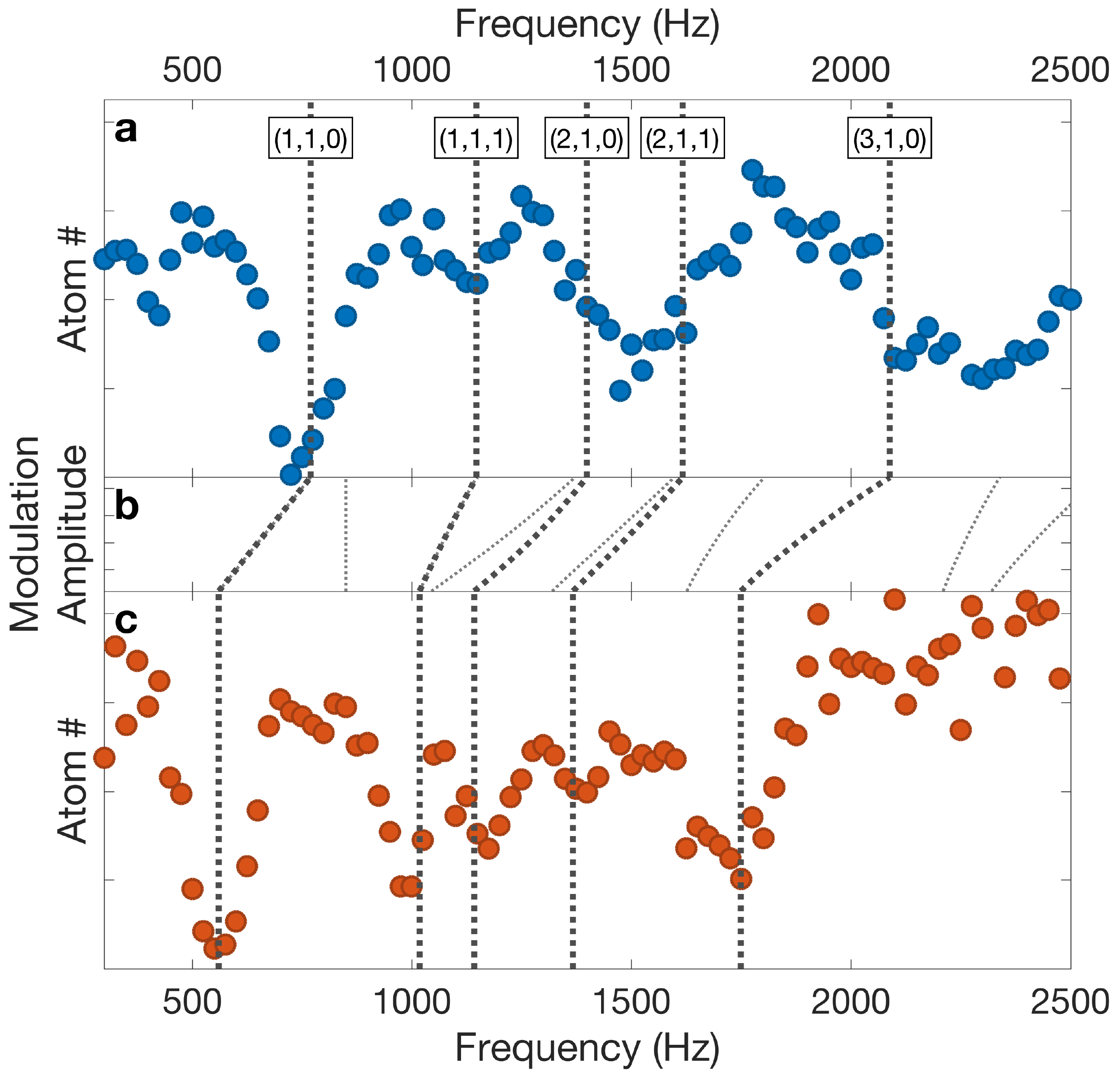}
\caption{\textbf{Tunable excitation spectra via trap shaping.}  \textbf{a:}~Bound fraction after a 1-second pulse as a function of excitation frequency for an unmodified trap. Bold dotted lines are theoretically predicted frequencies of collective resonances expected to couple to our drive. Labels on theory lines indicate the quantum numbers $(k,\beta,\gamma)$, using the notation of~\cite{graham-anisotropictrapexc}. $\beta$ and $\gamma$ are parity quantum numbers, and $k$ indicates the form of the nodal surface for the excitation. The quantum number $m$ is 1 for all resonances plotted. The only inputs to this theory are the three trap frequencies.   The resonance at half the fundamental frequency is believed to be due to parametric excitation of a dipole oscillation in the direction of gravity. Pulse amplitudes were increased from 0.6 $\upmu$m at low frequency to 3 $\upmu$m at the highest frequency to maximize peak visibility. \textbf{b:}~Evolution of predicted resonances under continuously increasing trap broadening. Thinner dotted lines represent resonances which are not dipole-allowed for this drive polarization.  \textbf{c:}~Bound fraction after a 1-second pulse as a function of excitation frequency for a trap broadened in one direction as described in the methods section. }
\label{fig:broadening}
\end{figure}

The excitation spectrum can be tuned by adjusting the trap shape, enabling the study of ultrafast-equivalent dynamics in systems with specific spectral characteristics such as mode degeneracies. The results of such tuning of the excitation spectrum are presented in Fig.~\ref{fig:broadening}.  We observe good agreement with analytic predictions for dipole-allowed collective resonance positions in the broadened and unbroadened trap~\cite{graham-anisotropictrapexc}.  Note that the frequencies of these complex anharmonic modes are not simply rescaled by broadening, but disperse at different rates; this enables tunable creation of mode degeneracies. This tunability of the collective excitation spectra is a key feature of cold-atom based quantum simulation of ultrafast dynamics. Static adjustments like those demonstrated here enable the realization of desired spectral properties, and rapid tuning of mode degeneracies could enable the study of controllably diabatic or adiabatic dynamics. Future experiments could use this ability for quantum simulation of molecular energy relaxation mechanisms in the vicinity of tunable mode degeneracies similar to conical intersections~\cite{CI-2001}.

\begin{figure}[t!]
\centering
\includegraphics[width=0.95\linewidth]{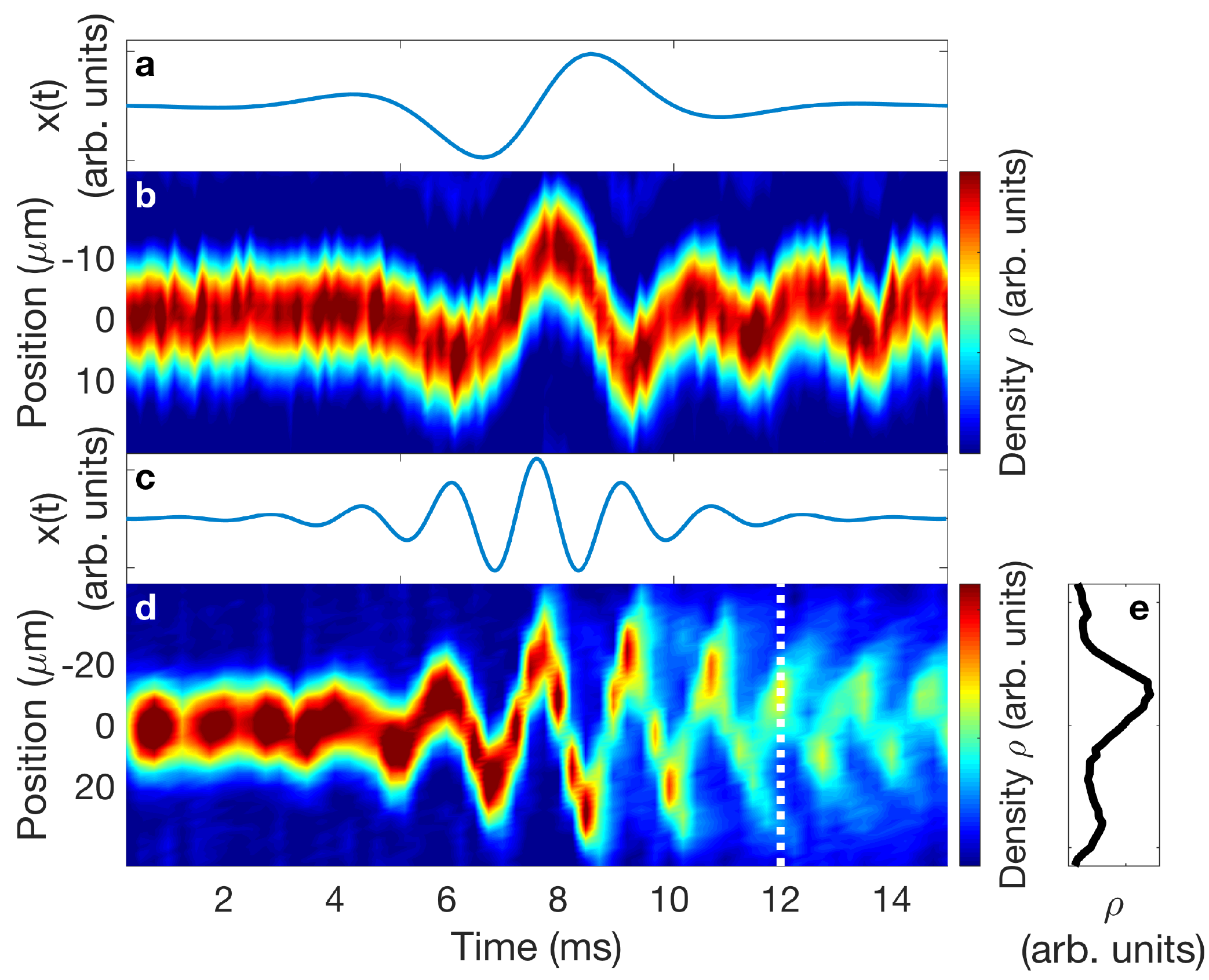}
\caption{\textbf{Sub-cycle dynamics during off-resonant and near-resonant pulses.} \textbf{a:}~Trap minimum position as a function of time for a pulse with $\tau = 3.76$~ms, $f = 200$~Hz, $A = 3 \upmu$m and $\phi = \mathrm{\pi}$. \textbf{b:}~Post-time-of-flight integrated spatial density distribution versus trap turn-off time during the off-resonant pulse depicted in panel \textbf{a}.  Here $\nu_{x}=450$~Hz. \textbf{c:}~Trap minimum position as a function of time for a pulse with $\tau = 3.76$~ms, $f = 550$~Hz, $A = 1.5 \upmu$m and $\phi = \frac{3 \mathrm{\pi}}{2}$. \textbf{d:}~Post-time-of-flight integrated spatial density distribution versus trap turn-off time during the near-resonant pulse depicted in panel \textbf{c}. Here $\nu_{x}=600$~Hz. \textbf{e:}~Density distribution at the time indicated by the dashed line in panel \textbf{d}.  Peaks from bound and ejected atoms are visible.}
\label{fig:streak}
\end{figure}

\begin{figure*}[t!]
\centering
\includegraphics[width=0.75\linewidth]{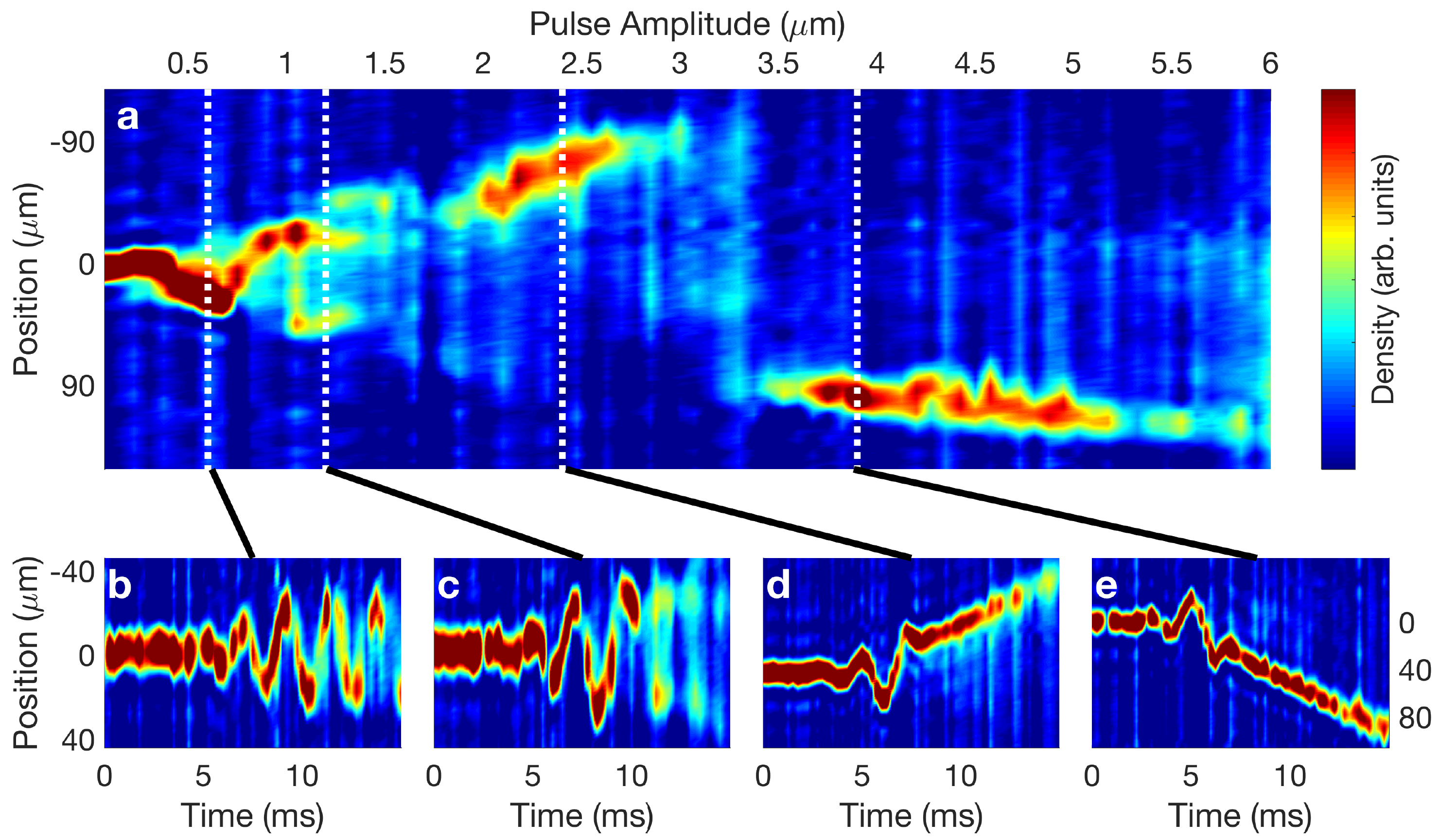}
\caption{\textbf{Dependence of unbinding dynamics on pulse amplitude.} \textbf{a:}~Integrated spatial density distribution after application of a near-resonant 480 Hz pulse with $\tau = 3.76$ ms and $\phi=0$ followed by 2 ms time-of-flight, versus pulse amplitude~$A$. \textbf{b-e:}~Integrated spatial density distribution versus time during pulses with the indicated amplitude. Panels \textbf{d} and \textbf{e} have an expanded $y$-axis (indicated at right) to track the unbound atoms. Note the momentum of the unbound atoms changing sign as the pulse amplitude increases.
}
\label{fig:amp}
\end{figure*}

\textbf{Momentum-resolved sub-cycle unbinding dynamics.} Having demonstrated quasi-CW spectroscopy of bound states with tunable energy spectra, we turn to the use of this tool for quantum simulation of ultrafast dynamics during few-cycle pulses~\cite{fewcycleionization-zerorange}. In ultrafast streaking measurements, the electric field of a few-cycle femtosecond pulse deflects photoelectrons produced by an attosecond extreme ultraviolet pulse striking an atom, allowing characterization of both the pulses and the atom~\cite{Hentschel2001,Itatani2002,Krausz2009}.  In the quantum simulator, qualitatively similar techniques allow high-resolution measurement of sub-cycle quantum dynamics. Here, instead of using photoionization to terminate the dynamics, the experimenter can simply instantaneously turn off the trapping potential at any point before, during or after the pulse. The atoms then propagate freely in space, and their instantaneous momenta at the time of trap removal are mapped onto their positions after some time of flight. Varying the time at which the trap is removed enables measurement of the time evolution of the bound quantum system with time resolution far below a drive period. This experimental technique, while commonplace in ultracold atomic physics, represents a powerful and general tool for the study of ultrafast-equivalent dynamics in our quantum simulator.

Fig.~\ref{fig:streak} presents the results of such measurements for both off-resonant and near-resonant pulses. The Bose-condensed atoms initially occupy mainly a single eigenstate of the transverse trapping potential. Quantum dynamics during and after the pulse can be tracked by direct momentum-space imaging of the atoms. Panels \textbf{b} and \textbf{d} of Fig.~\ref{fig:streak} show the density distribution after time-of-flight, integrated over the directions transverse to the excitation, as a function of time. For a pulse carrier frequency significantly below the dipole oscillation frequency $\nu_{x}$ in the dimension of driving, the momentum of the BEC evolves coherently during and after the pulse, as shown in Fig.~\ref{fig:streak}b. Incoherent heating due to the pulse is observed to be minimal on the few-cycle time scales we probe.   The atoms respond to the pulse at $\nu_{x}$ --- a higher frequency than the carrier --- but remain bound. During a pulse with carrier frequency near $\nu_{x}$, however, qualitatively different  is observed. Fig.~\ref{fig:streak}d shows momentum evolution during a near-resonant pulse for an amplitude near the unbinding threshold. In this parameter regime, atoms do not leave the trap all at once, but do not incoherently heat either; instead, ejection starts at the time of the pulse peak, with additional bursts of atoms emitted during each subsequent half-cycle of the pulse. Fig~\ref{fig:streak}e shows one such burst. This unbinding process models ionization or molecular disintegration during an ultrafast laser pulse.  

\textbf{Varying pulse amplitude and carrier-envelope phase.} The ability to precisely measure the population and momenta of unbound states as a function of pulse parameters and time opens up the possibility of flexible quantum simulation of ultrafast unbinding dynamics.  As an initial application of the quantum simulator presented herein we have measured the dependence of simulated ionization yield or photodissociation on both pulse amplitude and carrier-envelope phase. This represents a complementary method of testing the effects of two parameters central to numerous experimental and theoretical studies of ultrafast multiphoton ionization and bond-breaking processes~\cite{Keldysh1964,Faisal1973,Reiss1980,Chin1970,Mainfray1984,LHuillier1983}. 

Both the precise unbinding time during an applied force pulse and the final unbound momentum depend sensitively and non-monotonically on pulse amplitude. In the quantum simulator, the amplitude of the pulse can be straightforwardly varied over a wide range, keeping the carrier-envelope phase, carrier frequency, and total pulse time constant. As the amplitude is increased from that used in Fig.~\ref{fig:streak}d, the unbinding dynamics change drastically. Fig.~\ref{fig:amp}a shows the momentum distribution of the atoms (measured by detecting the position distribution after 2 ms time of flight) after few-cycle pulses with amplitudes from 0 up to 6~$\upmu$m. The bottom panels show the full time evolution of the momentum distribution during few-cycle pulses of selected amplitudes. Below a critical amplitude, no atoms are ejected from the trap. For some intermediate amplitudes, the behaviour mirrors that shown in Fig.~\ref{fig:streak}d, with bursts of atoms unbinding at different points during the pulse. Above that intermediate regime, all of the atoms  unbind at one well-defined time and continue to move with constant momentum after unbinding. Strikingly, as the amplitude is increased further, the momentum of the unbound atoms reverses sign, as they unbind half a drive cycle earlier, in an oppositely-directed simulated electric field.

\begin{figure}[t!]
\centering
\includegraphics[width=0.99\columnwidth]{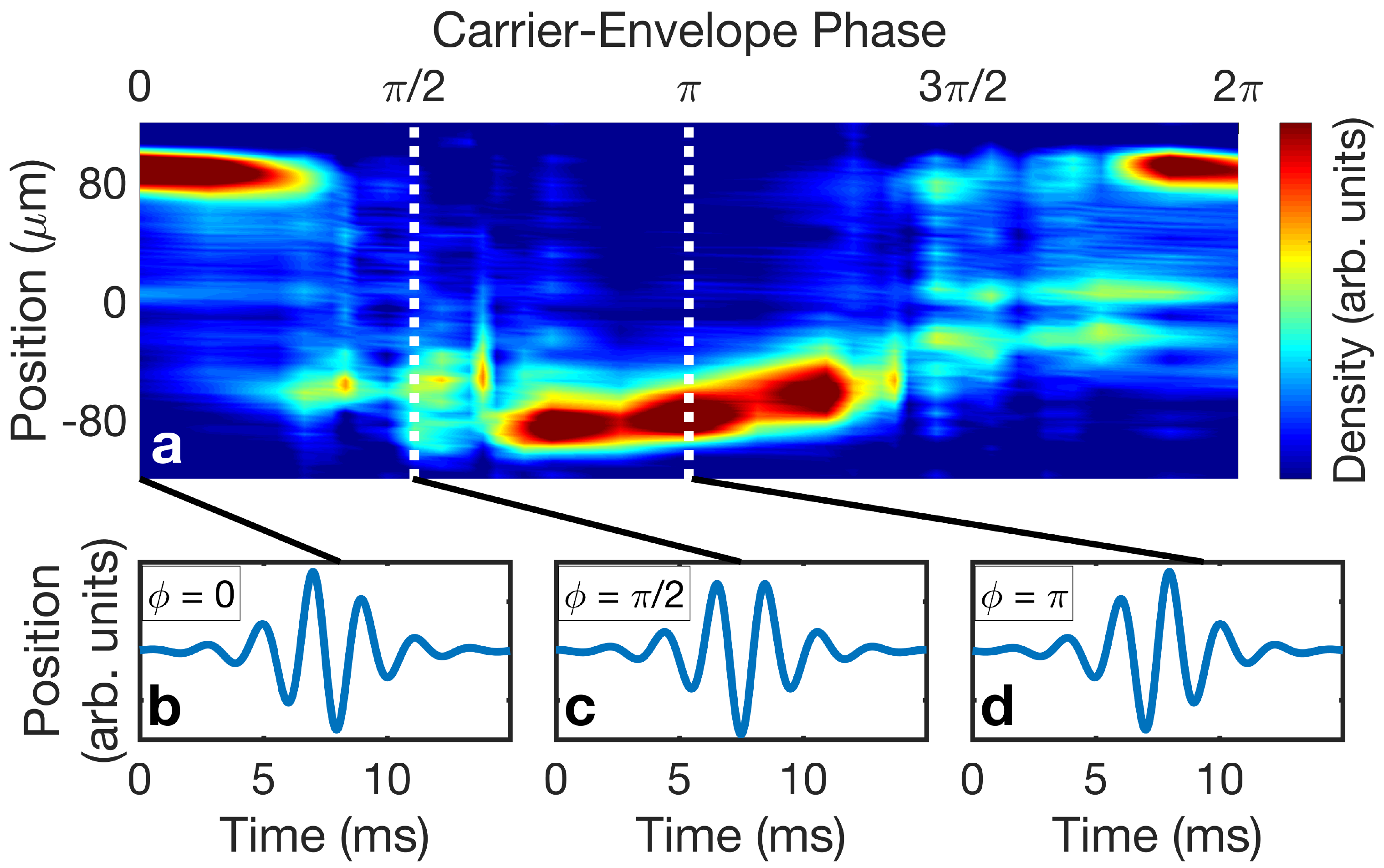}
\caption{\textbf{Carrier-envelope phase dependence of final momentum.} \textbf{a:} Integrated spatial density distribution after application of a near-resonant 450 Hz pulse with $\tau=3.76$ ms and $A=2.4$ $\upmu$m followed by 3 ms of time-of-flight, versus CEP. \textbf{b-d:} Pulse waveforms (force versus time) for CEP values indicated in the inset.}
\label{fig:CEP1}
\end{figure}

Even for fixed pulse amplitude, the final state of the initially bound system after the force pulse depends sensitively and non-trivially on the carrier-envelope phase $\phi$ (CEP).  In the pulsed-laser experimental context, the advent of few-cycle pulses with adjustable, stabilized CEP~\cite{Baltuska2003} has enabled advances such as probes of the effects of CEP on ultrafast dynamics~\cite{Peng2007,Kling2008}, study of interference patterns in multiparticle ionization signals~\cite{Kruger2011}, and control of recollision processes in molecular ions~\cite{Kling2006,Rathje2013}.
The nearly arbitrary pulse-shape control available in the cold-atom quantum simulator makes it a flexible tool for probing the dependence of ultrafast-equivalent dynamics on CEP. Fig.~\ref{fig:CEP1} shows a measurement of post-pulse momentum distribution (again detected via time of flight) as a function of the carrier-envelope phase of a near-resonant applied pulse. Changing the CEP from 0 to $\mathrm{\pi}$ flips the sign of all forces during the pulse, and results in inversion of the momentum of the unbound atoms.  As shown in the bottom panel of Fig.~\ref{fig:CEP1}, the pulses at integer values of $\phi/\mathrm{\pi}$ have sine-like character, possessing odd symmetry under reflection in time around the pulse centre.  Pulses with a CEP of $3\mathrm{\pi}/2$ have cosine-like character and give rise to very different unbinding dynamics at this pulse amplitude, populating more than one momentum class of unbound atoms. More complex dynamical phenomena are also visible in Fig.~\ref{fig:CEP1}: the inward slope of the unbound momentum as CEP increases in the neighborhood of $\phi=\mathrm{\pi}$ can be understood as the consequence of the force at the first unbinding peak sliding down the pulse envelope, and the observed asymmetry between $\phi=\mathrm{\pi}/2$ and $\phi=3\mathrm{\pi}/2$ indicates a violation of inversion symmetry in the potential. The most likely cause of this symmetry-breaking is slight trap aberration; this points the way to future work elucidating the effects of  potential shape on unbinding dynamics.

\section{Discussion}
The results presented here open the door to a broad class of quantum simulation experiments investigating ultrafast nonequilibrium phenomena, with numerous possible scientific targets. Emulation of pump-probe experiments, multichromatic light fields, and non-physical (for example, half-cycle) pulse shapes impossible to create with lasers would require no techniques beyond those demonstrated here apart from changing the form of $\alpha(t)$. 

Additional possibilities require only modest extensions of the experimental approach reported here. The simplest such extension would be to replace the inertial forces used to emulate electric fields with time-varying Zeeman or Stark potential gradients.  This would allow ultrafast quantum simulation in the regime of Keldysh parameter $\gamma_\mathrm{K}$ less than one, and enable the direct experimental investigation of open questions of current interest in ultrafast science. Potential scientific targets include the creation of photoelectron vortices with circularly-polarized pulses~\cite{starace-polarization}, the demonstration and study of strong-field stabilization, wherein the ionization probability becomes a decreasing function of pulse amplitude~\cite{stronfieldstabilization}, and detailed quantitative measurement of sub-cycle tunnel ionization timing effects~\cite{dudovich-tunnelingtime,attoclock-tunnelingtiming,krausz-tunnelingtime,lindrothhuillierphotoionization}. 

Other extensions to the basic technique are also possible. The use of  traps with multiple minima could enable modelling of more complex molecular configurations~\cite{sengstock-moleculesimulator}.  Ultrafast quantum simulation could also be pursued with small numbers of trapped fermions~\cite{jochim-fewfermion}, making a more direct analogue of atomic electrons.  However, the use of Bose condensates and the analogy of Eq.~\ref{GPE} does greatly magnify the signal, making experiments feasible with bosons that would be very challenging with fermions. Finally, an expansion of the analogy underlying these quantum simulation experiments beyond atoms and molecules could  enable the study of ultrafast-equivalent dynamical phenomena relevant to nuclear excitations~\cite{zinner-nuclearQS} and strong-field dynamics in solids~\cite{holthaus-strongfieldsim}.

In summary, we present experimental results from a cold-atom quantum simulator of ultrafast phenomena, including nonlinear spectroscopy of collective excitations, control of the energy spectra of the bound states of the simulator, imaging of sub-cycle dynamics during an unbinding process similar to ultrafast ionization, and measurement of the effects on unbinding dynamics of pulse amplitude and carrier-envelope phase. Such cold atom quantum simulation of ultrafast dynamical phenomena has the potential to enable benchmarking of relevant theories and explorations of experimentally challenging regimes, in an approach complementary to both ultrafast theory and pulsed-laser experiments. 

\section{Methods}
\textbf{Preparation of a Bose-Einstein condensate in an optical trap.} During the computer-controlled experimental sequence~\cite{Cicero}, atoms from an effusive source are collimated by a nozzle~\cite{nozzleRSI}, Zeeman-slowed, trapped and pre-cooled by sequential magneto-optical traps using the 461 nm and 689 nm ground-state transitions, and evaporatively cooled to degeneracy in a crossed-beam 1064 nm optical dipole trap (ODT)~\cite{strontium84BEC}. In the emulation stage of the experiment, the resulting condensate of $2\times10^4$ $^{84}$Sr atoms is adiabatically loaded into a single-beam ODT with a waist of 15~$\upmu$m. The trap depth can be varied across a wide range; a typical value used in the work presented here is 10~$\upmu$K. The s-wave scattering length of $^{84}$Sr is 6.5~nm.

\textbf{Temporal magnification of the quantum simulator.} Varying the power and shape of the trap beam yields transverse trap frequencies $\nu_x$ and $\nu_z$ between $300$ and $1000$ Hz, and an axial trap frequency $\nu_y$ between 5 and 15 Hz. Drawing the analogy between the optical trap and a single hydrogen atom, the energy difference  $h\nu_x$ between the ground and first relevant excited states (typically $\simeq$500 Hz) is analogous to the 2.47 PHz Ly-$\alpha$ line. Although the detailed energy spectrum of the ODT differs from that of hydrogen due to the different potential shapes, comparison of these two frequency scales indicates an approximate temporal magnification factor of 5$\times 10^{12}$. Emulation of molecular vibrational excitations or cluster dynamics leads to a  ratio of up to 10$^{11}$ between characteristic timescales of the emulator and emulated system. 

\textbf{Application of time-varying inertial forces.} An acousto-optic modulator (AOM) can translate the trap centre in the $x$-direction (see Fig.~\ref{fig:summary}c for axis definitions) at amplitudes up to 6~$\upmu$m and frequencies from DC up to hundreds of kHz. Because of the direction of translation and the wide separation between transverse and axial frequency scales, the axial degree of freedom is irrelevant to the results we present. Under the assumption that the atoms remain near the centre of the trap during the pulse, which we observe to be true until unbinding, the effective applied force is \mbox{$m\ddot{x} = F(t) = -dV(x,t)/dx\vert_{x=0}$}. $\ddot{x}$ can be specified as desired; for the particular functional form chosen, $\ddot{x}(t)$ has a similar enveloped-pulse shape to $x(t)$.   

\textbf{Tuning the trap geometry.} Shaping of the trap is achieved most simply by periodic translation on much faster time scales than the dynamics of the BEC, so that the atoms experience a time-averaged potential.  For this purpose, the trap AOM's RF drive frequency was sinusoidally modulated at 500~kHz, giving rise to a maximum trap translation amplitude of 4.2$\pm$0.3~$\upmu$m.

\textbf{Data availability.} The data that support the findings in this study are available from the corresponding author upon reasonable request.

\begin{acknowledgments}
The authors thank Vyacheslav Lebedev, Mikhail Lipatov, Ethan Q.\ Simmons, Yi Zeng, Jacob Hines, Ian Harley-Trochimczyk, James Chow,  and Bryance Oyang for experimental assistance, thank Andrew Jayich and Alejandro Saenz for critical readings of the manuscript, and acknowledge support from the National Science Foundation (CAREER 1555313), 
Air Force Office of Scientific Research (YIP FA9550-12-1-0305), Army Research Office (PECASE W911NF1410154 and DURIP W911NF1510436), and a President's Research Catalyst Award (CA-15-327861) from the UC Office of the President. 
\end{acknowledgments}

\section*{author contributions}
All authors contributed extensively to the work presented in this paper. RS, SVR, TS, and PED performed the measurements, with experimental assistance from KMF, KS, and ZAG. RS, SVR, TS, PED and DMW analysed the data and wrote the manuscript. All authors contributed insights and discussions on the experiment. DMW supervised the project.

\section*{Additional Information}
\textbf{Competing financial interests:} The authors declare no competing financial interests.


\begin{thebibliography}{54}%
\makeatletter
\providecommand \@ifxundefined [1]{%
 \@ifx{#1\undefined}
}%
\providecommand \@ifnum [1]{%
 \ifnum #1\expandafter \@firstoftwo
 \else \expandafter \@secondoftwo
 \fi
}%
\providecommand \@ifx [1]{%
 \ifx #1\expandafter \@firstoftwo
 \else \expandafter \@secondoftwo
 \fi
}%
\providecommand \natexlab [1]{#1}%
\providecommand \enquote  [1]{``#1''}%
\providecommand \bibnamefont  [1]{#1}%
\providecommand \bibfnamefont [1]{#1}%
\providecommand \citenamefont [1]{#1}%
\providecommand \href@noop [0]{\@secondoftwo}%
\providecommand \href [0]{\begingroup \@sanitize@url \@href}%
\providecommand \@href[1]{\@@startlink{#1}\@@href}%
\providecommand \@@href[1]{\endgroup#1\@@endlink}%
\providecommand \@sanitize@url [0]{\catcode `\\12\catcode `\$12\catcode
  `\&12\catcode `\#12\catcode `\^12\catcode `\_12\catcode `\%12\relax}%
\providecommand \@@startlink[1]{}%
\providecommand \@@endlink[0]{}%
\providecommand \url  [0]{\begingroup\@sanitize@url \@url }%
\providecommand \@url [1]{\endgroup\@href {#1}{\urlprefix }}%
\providecommand \urlprefix  [0]{URL }%
\providecommand \Eprint [0]{\href }%
\providecommand \doibase [0]{http://dx.doi.org/}%
\providecommand \selectlanguage [0]{\@gobble}%
\providecommand \bibinfo  [0]{\@secondoftwo}%
\providecommand \bibfield  [0]{\@secondoftwo}%
\providecommand \translation [1]{[#1]}%
\providecommand \BibitemOpen [0]{}%
\providecommand \bibitemStop [0]{}%
\providecommand \bibitemNoStop [0]{.\EOS\space}%
\providecommand \EOS [0]{\spacefactor3000\relax}%
\providecommand \BibitemShut  [1]{\csname bibitem#1\endcsname}%
\let\auto@bib@innerbib\@empty
\bibitem [{\citenamefont {Dum}\ \emph {et~al.}(1998)\citenamefont {Dum},
  \citenamefont {Sanpera}, \citenamefont {Suominen}, \citenamefont {Brewczyk},
  \citenamefont {Ku\ifmmode~\acute{s}\else \'{s}\fi{}}, \citenamefont
  {Rza\ifmmode \mbox{\c{}}\else \c{}\fi{}z\ifmmode~\dot{}\else
  \.{}\fi{}ewski},\ and\ \citenamefont
  {Lewenstein}}]{lewenstein-wavepacketdynamics}%
  \BibitemOpen
  \bibfield  {author} {\bibinfo {author} {\bibfnamefont {R.}~\bibnamefont
  {Dum}}, \bibinfo {author} {\bibfnamefont {A.}~\bibnamefont {Sanpera}},
  \bibinfo {author} {\bibfnamefont {K.-A.}\ \bibnamefont {Suominen}}, \bibinfo
  {author} {\bibfnamefont {M.}~\bibnamefont {Brewczyk}}, \bibinfo {author}
  {\bibfnamefont {M.}~\bibnamefont {Ku\ifmmode~\acute{s}\else \'{s}\fi{}}},
  \bibinfo {author} {\bibfnamefont {K.}~\bibnamefont {Rza\ifmmode
  \mbox{\c{}}\else \c{}\fi{}z\ifmmode~\dot{}\else \.{}\fi{}ewski}}, \ and\
  \bibinfo {author} {\bibfnamefont {M.}~\bibnamefont {Lewenstein}},\ }\emph
  {\bibinfo {title} {Wave Packet Dynamics with Bose-Einstein Condensates}},\
  \href {\doibase 10.1103/PhysRevLett.80.3899} {\bibfield  {journal} {\bibinfo
  {journal} {Phys. Rev. Lett.}\ }\textbf {\bibinfo {volume} {80}}, \bibinfo
  {pages} {3899--3902} (\bibinfo {year} {1998})}\BibitemShut {NoStop}%
\bibitem [{\citenamefont {Sala}\ \emph {et~al.}(2017)\citenamefont {Sala},
  \citenamefont {F\"orster},\ and\ \citenamefont
  {Saenz}}]{saenz-attosecondsimulatorPRA}%
  \BibitemOpen
  \bibfield  {author} {\bibinfo {author} {\bibfnamefont {S.}~\bibnamefont
  {Sala}}, \bibinfo {author} {\bibfnamefont {J.}~\bibnamefont {F\"orster}}, \
  and\ \bibinfo {author} {\bibfnamefont {A.}~\bibnamefont {Saenz}},\ }\emph
  {\bibinfo {title} {Ultracold-atom quantum simulator for attosecond
  science}},\ \href {\doibase 10.1103/PhysRevA.95.011403} {\bibfield  {journal}
  {\bibinfo  {journal} {Phys. Rev. A}\ }\textbf {\bibinfo {volume} {95}},
  \bibinfo {pages} {011403} (\bibinfo {year} {2017})}\BibitemShut {NoStop}%
\bibitem [{\citenamefont {Arlinghaus}\ and\ \citenamefont
  {Holthaus}(2010)}]{holthaus-strongfieldsim}%
  \BibitemOpen
  \bibfield  {author} {\bibinfo {author} {\bibfnamefont {S.}~\bibnamefont
  {Arlinghaus}}\ and\ \bibinfo {author} {\bibfnamefont {M.}~\bibnamefont
  {Holthaus}},\ }\emph {\bibinfo {title} {Driven optical lattices as
  strong-field simulators}},\ \href {\doibase 10.1103/PhysRevA.81.063612}
  {\bibfield  {journal} {\bibinfo  {journal} {Phys. Rev. A}\ }\textbf {\bibinfo
  {volume} {81}}, \bibinfo {pages} {063612} (\bibinfo {year}
  {2010})}\BibitemShut {NoStop}%
\bibitem [{\citenamefont {L\"uhmann}\ \emph {et~al.}(2015)\citenamefont
  {L\"uhmann}, \citenamefont {Weitenberg},\ and\ \citenamefont
  {Sengstock}}]{sengstock-moleculesimulator}%
  \BibitemOpen
  \bibfield  {author} {\bibinfo {author} {\bibfnamefont {D.-S.}\ \bibnamefont
  {L\"uhmann}}, \bibinfo {author} {\bibfnamefont {C.}~\bibnamefont
  {Weitenberg}}, \ and\ \bibinfo {author} {\bibfnamefont {K.}~\bibnamefont
  {Sengstock}},\ }\emph {\bibinfo {title} {Emulating Molecular Orbitals and
  Electronic Dynamics with Ultracold Atoms}},\ \href {\doibase
  10.1103/PhysRevX.5.031016} {\bibfield  {journal} {\bibinfo  {journal} {Phys.
  Rev. X}\ }\textbf {\bibinfo {volume} {5}}, \bibinfo {pages} {031016}
  (\bibinfo {year} {2015})}\BibitemShut {NoStop}%
\bibitem [{\citenamefont {Rajagopal}\ \emph {et~al.}(2017)\citenamefont
  {Rajagopal}, \citenamefont {Fujiwara}, \citenamefont {Senaratne},
  \citenamefont {Singh}, \citenamefont {Geiger},\ and\ \citenamefont
  {Weld}}]{adppaper}%
  \BibitemOpen
  \bibfield  {author} {\bibinfo {author} {\bibfnamefont {S.~V.}\ \bibnamefont
  {Rajagopal}}, \bibinfo {author} {\bibfnamefont {K.~M.}\ \bibnamefont
  {Fujiwara}}, \bibinfo {author} {\bibfnamefont {R.}~\bibnamefont {Senaratne}},
  \bibinfo {author} {\bibfnamefont {K.}~\bibnamefont {Singh}}, \bibinfo
  {author} {\bibfnamefont {Z.~A.}\ \bibnamefont {Geiger}}, \ and\ \bibinfo
  {author} {\bibfnamefont {D.~M.}\ \bibnamefont {Weld}},\ }\emph {\bibinfo
  {title} {Quantum Emulation of Extreme Non-Equilibrium Phenomena with Trapped
  Atoms}},\ \href {\doibase 10.1002/andp.201700008} {\bibfield  {journal}
  {\bibinfo  {journal} {Ann. Phys.}\ }\textbf {\bibinfo {volume} {529}},
  \bibinfo {pages} {1700008} (\bibinfo {year} {2017})}\BibitemShut {NoStop}%
\bibitem [{\citenamefont {Lewenstein}\ \emph {et~al.}(2007)\citenamefont
  {Lewenstein}, \citenamefont {Sanpera}, \citenamefont {Ahufinger},
  \citenamefont {Damski}, \citenamefont {Sen},\ and\ \citenamefont
  {Sen}}]{lewenstein-review}%
  \BibitemOpen
  \bibfield  {author} {\bibinfo {author} {\bibfnamefont {M.}~\bibnamefont
  {Lewenstein}}, \bibinfo {author} {\bibfnamefont {A.}~\bibnamefont {Sanpera}},
  \bibinfo {author} {\bibfnamefont {V.}~\bibnamefont {Ahufinger}}, \bibinfo
  {author} {\bibfnamefont {B.}~\bibnamefont {Damski}}, \bibinfo {author}
  {\bibfnamefont {A.}~\bibnamefont {Sen}}, \ and\ \bibinfo {author}
  {\bibfnamefont {U.}~\bibnamefont {Sen}},\ }\emph {\bibinfo {title} {Ultracold
  atomic gases in optical lattices: mimicking condensed matter physics and
  beyond}},\ \href {\doibase 10.1080/00018730701223200} {\bibfield  {journal}
  {\bibinfo  {journal} {Adv. Phys.}\ }\textbf {\bibinfo {volume} {56}},
  \bibinfo {pages} {243--379} (\bibinfo {year} {2007})}\BibitemShut {NoStop}%
\bibitem [{\citenamefont {Bloch}\ \emph {et~al.}(2012)\citenamefont {Bloch},
  \citenamefont {Dalibard},\ and\ \citenamefont
  {Nascimbene}}]{blochdalibardQSreview}%
  \BibitemOpen
  \bibfield  {author} {\bibinfo {author} {\bibfnamefont {I.}~\bibnamefont
  {Bloch}}, \bibinfo {author} {\bibfnamefont {J.}~\bibnamefont {Dalibard}}, \
  and\ \bibinfo {author} {\bibfnamefont {S.}~\bibnamefont {Nascimbene}},\
  }\emph {\bibinfo {title} {Quantum simulations with ultracold quantum
  gases}},\ \href {http://dx.doi.org/10.1038/nphys2259} {\bibfield  {journal}
  {\bibinfo  {journal} {Nat. Phys.}\ }\textbf {\bibinfo {volume} {8}}, \bibinfo
  {pages} {267--276} (\bibinfo {year} {2012})}\BibitemShut {NoStop}%
\bibitem [{\citenamefont {Gross}\ and\ \citenamefont
  {Bloch}(2017)}]{bloch2017review}%
  \BibitemOpen
  \bibfield  {author} {\bibinfo {author} {\bibfnamefont {C.}~\bibnamefont
  {Gross}}\ and\ \bibinfo {author} {\bibfnamefont {I.}~\bibnamefont {Bloch}},\
  }\emph {\bibinfo {title} {Quantum simulations with ultracold atoms in optical
  lattices}},\ \href {\doibase 10.1126/science.aal3837} {\bibfield  {journal}
  {\bibinfo  {journal} {Science}\ }\textbf {\bibinfo {volume} {357}}, \bibinfo
  {pages} {995--1001} (\bibinfo {year} {2017})}\BibitemShut {NoStop}%
\bibitem [{\citenamefont {Greiner}\ \emph {et~al.}(2002)\citenamefont
  {Greiner}, \citenamefont {Mandel}, \citenamefont {Esslinger}, \citenamefont
  {H{\"a}nsch},\ and\ \citenamefont {Bloch}}]{greinerSFMI}%
  \BibitemOpen
  \bibfield  {author} {\bibinfo {author} {\bibfnamefont {M.}~\bibnamefont
  {Greiner}}, \bibinfo {author} {\bibfnamefont {O.}~\bibnamefont {Mandel}},
  \bibinfo {author} {\bibfnamefont {T.}~\bibnamefont {Esslinger}}, \bibinfo
  {author} {\bibfnamefont {T.~W.}\ \bibnamefont {H{\"a}nsch}}, \ and\ \bibinfo
  {author} {\bibfnamefont {I.}~\bibnamefont {Bloch}},\ }\emph {\bibinfo {title}
  {Quantum phase transition from a superfluid to a Mott insulator in a gas of
  ultracold atoms}},\ \href {\doibase 10.1038/415039a} {\bibfield  {journal}
  {\bibinfo  {journal} {Nature}\ }\textbf {\bibinfo {volume} {415}}, \bibinfo
  {pages} {39--44} (\bibinfo {year} {2002})}\BibitemShut {NoStop}%
\bibitem [{\citenamefont {Hart}\ \emph {et~al.}(2015)\citenamefont {Hart},
  \citenamefont {Duarte}, \citenamefont {Yang}, \citenamefont {Liu},
  \citenamefont {Paiva}, \citenamefont {Khatami}, \citenamefont {Scalettar},
  \citenamefont {Trivedi}, \citenamefont {Huse},\ and\ \citenamefont
  {Hulet}}]{HuletAF}%
  \BibitemOpen
  \bibfield  {author} {\bibinfo {author} {\bibfnamefont {R.~A.}\ \bibnamefont
  {Hart}}, \bibinfo {author} {\bibfnamefont {P.~M.}\ \bibnamefont {Duarte}},
  \bibinfo {author} {\bibfnamefont {T.-L.}\ \bibnamefont {Yang}}, \bibinfo
  {author} {\bibfnamefont {X.}~\bibnamefont {Liu}}, \bibinfo {author}
  {\bibfnamefont {T.}~\bibnamefont {Paiva}}, \bibinfo {author} {\bibfnamefont
  {E.}~\bibnamefont {Khatami}}, \bibinfo {author} {\bibfnamefont {R.~T.}\
  \bibnamefont {Scalettar}}, \bibinfo {author} {\bibfnamefont {N.}~\bibnamefont
  {Trivedi}}, \bibinfo {author} {\bibfnamefont {D.~A.}\ \bibnamefont {Huse}}, \
  and\ \bibinfo {author} {\bibfnamefont {R.~G.}\ \bibnamefont {Hulet}},\ }\emph
  {\bibinfo {title} {Observation of antiferromagnetic correlations in the
  Hubbard model with ultracold atoms}},\ \href
  {http://dx.doi.org/10.1038/nature14223} {\bibfield  {journal} {\bibinfo
  {journal} {Nature (London)}\ }\textbf {\bibinfo {volume} {519}}, \bibinfo
  {pages} {211--214} (\bibinfo {year} {2015})}\BibitemShut {NoStop}%
\bibitem [{\citenamefont {Ben~Dahan}\ \emph {et~al.}(1996)\citenamefont
  {Ben~Dahan}, \citenamefont {Peik}, \citenamefont {Reichel}, \citenamefont
  {Castin},\ and\ \citenamefont {Salomon}}]{salomon-blochoscs}%
  \BibitemOpen
  \bibfield  {author} {\bibinfo {author} {\bibfnamefont {M.}~\bibnamefont
  {Ben~Dahan}}, \bibinfo {author} {\bibfnamefont {E.}~\bibnamefont {Peik}},
  \bibinfo {author} {\bibfnamefont {J.}~\bibnamefont {Reichel}}, \bibinfo
  {author} {\bibfnamefont {Y.}~\bibnamefont {Castin}}, \ and\ \bibinfo {author}
  {\bibfnamefont {C.}~\bibnamefont {Salomon}},\ }\emph {\bibinfo {title} {Bloch
  Oscillations of Atoms in an Optical Potential}},\ \href {\doibase
  10.1103/PhysRevLett.76.4508} {\bibfield  {journal} {\bibinfo  {journal}
  {Phys. Rev. Lett.}\ }\textbf {\bibinfo {volume} {76}}, \bibinfo {pages}
  {4508--4511} (\bibinfo {year} {1996})}\BibitemShut {NoStop}%
\bibitem [{\citenamefont {Schreiber}\ \emph {et~al.}(2015)\citenamefont
  {Schreiber}, \citenamefont {Hodgman}, \citenamefont {Bordia}, \citenamefont
  {L{\"u}schen}, \citenamefont {Fischer}, \citenamefont {Vosk}, \citenamefont
  {Altman}, \citenamefont {Schneider},\ and\ \citenamefont
  {Bloch}}]{bloch-mblscience}%
  \BibitemOpen
  \bibfield  {author} {\bibinfo {author} {\bibfnamefont {M.}~\bibnamefont
  {Schreiber}}, \bibinfo {author} {\bibfnamefont {S.~S.}\ \bibnamefont
  {Hodgman}}, \bibinfo {author} {\bibfnamefont {P.}~\bibnamefont {Bordia}},
  \bibinfo {author} {\bibfnamefont {H.~P.}\ \bibnamefont {L{\"u}schen}},
  \bibinfo {author} {\bibfnamefont {M.~H.}\ \bibnamefont {Fischer}}, \bibinfo
  {author} {\bibfnamefont {R.}~\bibnamefont {Vosk}}, \bibinfo {author}
  {\bibfnamefont {E.}~\bibnamefont {Altman}}, \bibinfo {author} {\bibfnamefont
  {U.}~\bibnamefont {Schneider}}, \ and\ \bibinfo {author} {\bibfnamefont
  {I.}~\bibnamefont {Bloch}},\ }\emph {\bibinfo {title} {Observation of
  many-body localization of interacting fermions in a quasirandom optical
  lattice}},\ \href {\doibase 10.1126/science.aaa7432} {\bibfield  {journal}
  {\bibinfo  {journal} {Science}\ }\textbf {\bibinfo {volume} {349}}, \bibinfo
  {pages} {842--845} (\bibinfo {year} {2015})}\BibitemShut {NoStop}%
\bibitem [{\citenamefont {Krausz}\ and\ \citenamefont
  {Ivanov}(2009{\natexlab{a}})}]{krausz-attosecond-review}%
  \BibitemOpen
  \bibfield  {author} {\bibinfo {author} {\bibfnamefont {F.}~\bibnamefont
  {Krausz}}\ and\ \bibinfo {author} {\bibfnamefont {M.}~\bibnamefont
  {Ivanov}},\ }\emph {\bibinfo {title} {Attosecond physics}},\ \href {\doibase
  10.1103/RevModPhys.81.163} {\bibfield  {journal} {\bibinfo  {journal} {Rev.
  Mod. Phys.}\ }\textbf {\bibinfo {volume} {81}}, \bibinfo {pages} {163--234}
  (\bibinfo {year} {2009}{\natexlab{a}})}\BibitemShut {NoStop}%
\bibitem [{\citenamefont {Corkum}\ and\ \citenamefont
  {Krausz}(2007)}]{corkumkrauszreview}%
  \BibitemOpen
  \bibfield  {author} {\bibinfo {author} {\bibfnamefont {P.~B.}\ \bibnamefont
  {Corkum}}\ and\ \bibinfo {author} {\bibfnamefont {F.}~\bibnamefont
  {Krausz}},\ }\emph {\bibinfo {title} {Attosecond science}},\ \href
  {http://dx.doi.org/10.1038/nphys620} {\bibfield  {journal} {\bibinfo
  {journal} {Nat. Phys.}\ }\textbf {\bibinfo {volume} {3}}, \bibinfo {pages}
  {381--387} (\bibinfo {year} {2007})}\BibitemShut {NoStop}%
\bibitem [{\citenamefont {Fennel}\ \emph {et~al.}(2010)\citenamefont {Fennel},
  \citenamefont {Meiwes-Broer}, \citenamefont {Tiggesb\"aumker}, \citenamefont
  {Reinhard}, \citenamefont {Dinh},\ and\ \citenamefont {Suraud}}]{clusters}%
  \BibitemOpen
  \bibfield  {author} {\bibinfo {author} {\bibfnamefont {T.}~\bibnamefont
  {Fennel}}, \bibinfo {author} {\bibfnamefont {K.-H.}\ \bibnamefont
  {Meiwes-Broer}}, \bibinfo {author} {\bibfnamefont {J.}~\bibnamefont
  {Tiggesb\"aumker}}, \bibinfo {author} {\bibfnamefont {P.-G.}\ \bibnamefont
  {Reinhard}}, \bibinfo {author} {\bibfnamefont {P.~M.}\ \bibnamefont {Dinh}},
  \ and\ \bibinfo {author} {\bibfnamefont {E.}~\bibnamefont {Suraud}},\ }\emph
  {\bibinfo {title} {Laser-driven nonlinear cluster dynamics}},\ \href
  {\doibase 10.1103/RevModPhys.82.1793} {\bibfield  {journal} {\bibinfo
  {journal} {Rev. Mod. Phys.}\ }\textbf {\bibinfo {volume} {82}}, \bibinfo
  {pages} {1793--1842} (\bibinfo {year} {2010})}\BibitemShut {NoStop}%
\bibitem [{\citenamefont {Keldysh}(1964)}]{Keldysh1964}%
  \BibitemOpen
  \bibfield  {author} {\bibinfo {author} {\bibfnamefont {L.~V.}\ \bibnamefont
  {Keldysh}},\ }\emph {\bibinfo {title} {{IONIZATION IN THE FIELD OF A STRONG
  ELECTROMAGNETIC WAVE}}},\ \href
  {http://www.jetp.ac.ru/cgi-bin/dn/e{\_}020{\_}05{\_}1307.pdf} {\bibfield
  {journal} {\bibinfo  {journal} {J. Exp. Theor. Phys.}\ }\textbf {\bibinfo
  {volume} {47}}, \bibinfo {pages} {1945--1957} (\bibinfo {year}
  {1964})}\BibitemShut {NoStop}%
\bibitem [{\citenamefont {Faisal}(1973)}]{Faisal1973}%
  \BibitemOpen
  \bibfield  {author} {\bibinfo {author} {\bibfnamefont {F.~H.~M.}\
  \bibnamefont {Faisal}},\ }\emph {\bibinfo {title} {Multiphoton transitions.
  IV. Bound-free transition integrals in compact forms}},\ \href {\doibase
  10.1088/0022-3700/6/3/023} {\bibfield  {journal} {\bibinfo  {journal} {J.
  Phys. B: At. Mol. Phys.}\ }\textbf {\bibinfo {volume} {6}}, \bibinfo {pages}
  {553} (\bibinfo {year} {1973})}\BibitemShut {NoStop}%
\bibitem [{\citenamefont {Reiss}(1980)}]{Reiss1980}%
  \BibitemOpen
  \bibfield  {author} {\bibinfo {author} {\bibfnamefont {H.~R.}\ \bibnamefont
  {Reiss}},\ }\emph {\bibinfo {title} {Gauges for intense-field
  electrodynamics}},\ \href {\doibase 10.1103/PhysRevA.22.770} {\bibfield
  {journal} {\bibinfo  {journal} {Phys. Rev. A}\ }\textbf {\bibinfo {volume}
  {22}}, \bibinfo {pages} {770--772} (\bibinfo {year} {1980})}\BibitemShut
  {NoStop}%
\bibitem [{\citenamefont {Corkum}(1993)}]{corkum-model}%
  \BibitemOpen
  \bibfield  {author} {\bibinfo {author} {\bibfnamefont {P.~B.}\ \bibnamefont
  {Corkum}},\ }\emph {\bibinfo {title} {Plasma perspective on strong field
  multiphoton ionization}},\ \href {\doibase 10.1103/PhysRevLett.71.1994}
  {\bibfield  {journal} {\bibinfo  {journal} {Phys. Rev. Lett.}\ }\textbf
  {\bibinfo {volume} {71}}, \bibinfo {pages} {1994--1997} (\bibinfo {year}
  {1993})}\BibitemShut {NoStop}%
\bibitem [{\citenamefont {Lewenstein}\ \emph {et~al.}(1994)\citenamefont
  {Lewenstein}, \citenamefont {Balcou}, \citenamefont {Ivanov}, \citenamefont
  {L'Huillier},\ and\ \citenamefont {Corkum}}]{lewenstein-hhg}%
  \BibitemOpen
  \bibfield  {author} {\bibinfo {author} {\bibfnamefont {M.}~\bibnamefont
  {Lewenstein}}, \bibinfo {author} {\bibfnamefont {P.}~\bibnamefont {Balcou}},
  \bibinfo {author} {\bibfnamefont {M.~Y.}\ \bibnamefont {Ivanov}}, \bibinfo
  {author} {\bibfnamefont {A.}~\bibnamefont {L'Huillier}}, \ and\ \bibinfo
  {author} {\bibfnamefont {P.~B.}\ \bibnamefont {Corkum}},\ }\emph {\bibinfo
  {title} {Theory of high-harmonic generation by low-frequency laser fields}},\
  \href {\doibase 10.1103/PhysRevA.49.2117} {\bibfield  {journal} {\bibinfo
  {journal} {Phys. Rev. A}\ }\textbf {\bibinfo {volume} {49}}, \bibinfo {pages}
  {2117--2132} (\bibinfo {year} {1994})}\BibitemShut {NoStop}%
\bibitem [{\citenamefont {Jin}\ \emph {et~al.}(1996)\citenamefont {Jin},
  \citenamefont {Ensher}, \citenamefont {Matthews}, \citenamefont {Wieman},\
  and\ \citenamefont {Cornell}}]{jin-collectiveexc}%
  \BibitemOpen
  \bibfield  {author} {\bibinfo {author} {\bibfnamefont {D.~S.}\ \bibnamefont
  {Jin}}, \bibinfo {author} {\bibfnamefont {J.~R.}\ \bibnamefont {Ensher}},
  \bibinfo {author} {\bibfnamefont {M.~R.}\ \bibnamefont {Matthews}}, \bibinfo
  {author} {\bibfnamefont {C.~E.}\ \bibnamefont {Wieman}}, \ and\ \bibinfo
  {author} {\bibfnamefont {E.~A.}\ \bibnamefont {Cornell}},\ }\emph {\bibinfo
  {title} {Collective Excitations of a Bose-Einstein Condensate in a Dilute
  Gas}},\ \href {\doibase 10.1103/PhysRevLett.77.420} {\bibfield  {journal}
  {\bibinfo  {journal} {Phys. Rev. Lett.}\ }\textbf {\bibinfo {volume} {77}},
  \bibinfo {pages} {420--423} (\bibinfo {year} {1996})}\BibitemShut {NoStop}%
\bibitem [{\citenamefont {Mewes}\ \emph {et~al.}(1996)\citenamefont {Mewes},
  \citenamefont {Andrews}, \citenamefont {van Druten}, \citenamefont {Kurn},
  \citenamefont {Durfee}, \citenamefont {Townsend},\ and\ \citenamefont
  {Ketterle}}]{ketterle-collectiveexcitations}%
  \BibitemOpen
  \bibfield  {author} {\bibinfo {author} {\bibfnamefont {M.-O.}\ \bibnamefont
  {Mewes}}, \bibinfo {author} {\bibfnamefont {M.~R.}\ \bibnamefont {Andrews}},
  \bibinfo {author} {\bibfnamefont {N.~J.}\ \bibnamefont {van Druten}},
  \bibinfo {author} {\bibfnamefont {D.~M.}\ \bibnamefont {Kurn}}, \bibinfo
  {author} {\bibfnamefont {D.~S.}\ \bibnamefont {Durfee}}, \bibinfo {author}
  {\bibfnamefont {C.~G.}\ \bibnamefont {Townsend}}, \ and\ \bibinfo {author}
  {\bibfnamefont {W.}~\bibnamefont {Ketterle}},\ }\emph {\bibinfo {title}
  {Collective Excitations of a Bose-Einstein Condensate in a Magnetic Trap}},\
  \href {\doibase 10.1103/PhysRevLett.77.988} {\bibfield  {journal} {\bibinfo
  {journal} {Phys. Rev. Lett.}\ }\textbf {\bibinfo {volume} {77}}, \bibinfo
  {pages} {988--991} (\bibinfo {year} {1996})}\BibitemShut {NoStop}%
\bibitem [{\citenamefont {Fort}\ \emph {et~al.}(2000)\citenamefont {Fort},
  \citenamefont {Prevedelli}, \citenamefont {Minardi}, \citenamefont
  {Cataliotti}, \citenamefont {Ricci}, \citenamefont {Tino},\ and\
  \citenamefont {Inguscio}}]{inguscioexcitations}%
  \BibitemOpen
  \bibfield  {author} {\bibinfo {author} {\bibfnamefont {C.}~\bibnamefont
  {Fort}}, \bibinfo {author} {\bibfnamefont {M.}~\bibnamefont {Prevedelli}},
  \bibinfo {author} {\bibfnamefont {F.}~\bibnamefont {Minardi}}, \bibinfo
  {author} {\bibfnamefont {F.~S.}\ \bibnamefont {Cataliotti}}, \bibinfo
  {author} {\bibfnamefont {L.}~\bibnamefont {Ricci}}, \bibinfo {author}
  {\bibfnamefont {G.~M.}\ \bibnamefont {Tino}}, \ and\ \bibinfo {author}
  {\bibfnamefont {M.}~\bibnamefont {Inguscio}},\ }\emph {\bibinfo {title}
  {Collective excitations of a 87 Rb Bose condensate in the Thomas-Fermi
  regime}},\ \href {\doibase 10.1209/epl/i2000-00112-5} {\bibfield  {journal}
  {\bibinfo  {journal} {EPL (Europhysics Letters)}\ }\textbf {\bibinfo {volume}
  {49}}, \bibinfo {pages} {8} (\bibinfo {year} {2000})}\BibitemShut {NoStop}%
\bibitem [{\citenamefont {Stringari}(1996)}]{stringari-orig}%
  \BibitemOpen
  \bibfield  {author} {\bibinfo {author} {\bibfnamefont {S.}~\bibnamefont
  {Stringari}},\ }\emph {\bibinfo {title} {Collective Excitations of a Trapped
  Bose-Condensed Gas}},\ \href {\doibase 10.1103/PhysRevLett.77.2360}
  {\bibfield  {journal} {\bibinfo  {journal} {Phys. Rev. Lett.}\ }\textbf
  {\bibinfo {volume} {77}}, \bibinfo {pages} {2360--2363} (\bibinfo {year}
  {1996})}\BibitemShut {NoStop}%
\bibitem [{\citenamefont {\"Ohberg}\ \emph {et~al.}(1997)\citenamefont
  {\"Ohberg}, \citenamefont {Surkov}, \citenamefont {Tittonen}, \citenamefont
  {Stenholm}, \citenamefont {Wilkens},\ and\ \citenamefont
  {Shlyapnikov}}]{shlyapnikov-excitations}%
  \BibitemOpen
  \bibfield  {author} {\bibinfo {author} {\bibfnamefont {P.}~\bibnamefont
  {\"Ohberg}}, \bibinfo {author} {\bibfnamefont {E.~L.}\ \bibnamefont
  {Surkov}}, \bibinfo {author} {\bibfnamefont {I.}~\bibnamefont {Tittonen}},
  \bibinfo {author} {\bibfnamefont {S.}~\bibnamefont {Stenholm}}, \bibinfo
  {author} {\bibfnamefont {M.}~\bibnamefont {Wilkens}}, \ and\ \bibinfo
  {author} {\bibfnamefont {G.~V.}\ \bibnamefont {Shlyapnikov}},\ }\emph
  {\bibinfo {title} {Low-energy elementary excitations of a trapped
  Bose-condensed gas}},\ \href {\doibase 10.1103/PhysRevA.56.R3346} {\bibfield
  {journal} {\bibinfo  {journal} {Phys. Rev. A}\ }\textbf {\bibinfo {volume}
  {56}}, \bibinfo {pages} {R3346--R3349} (\bibinfo {year} {1997})}\BibitemShut
  {NoStop}%
\bibitem [{\citenamefont {Csord\'as}\ and\ \citenamefont
  {Graham}(1999)}]{graham-anisotropictrapexc}%
  \BibitemOpen
  \bibfield  {author} {\bibinfo {author} {\bibfnamefont {A.}~\bibnamefont
  {Csord\'as}}\ and\ \bibinfo {author} {\bibfnamefont {R.}~\bibnamefont
  {Graham}},\ }\emph {\bibinfo {title} {Collective excitations in Bose-Einstein
  condensates in triaxially anisotropic parabolic traps}},\ \href {\doibase
  10.1103/PhysRevA.59.1477} {\bibfield  {journal} {\bibinfo  {journal} {Phys.
  Rev. A}\ }\textbf {\bibinfo {volume} {59}}, \bibinfo {pages} {1477--1487}
  (\bibinfo {year} {1999})}\BibitemShut {NoStop}%
\bibitem [{\citenamefont {Esry}(1997)}]{esry97}%
  \BibitemOpen
  \bibfield  {author} {\bibinfo {author} {\bibfnamefont {B.~D.}\ \bibnamefont
  {Esry}},\ }\emph {\bibinfo {title} {Hartree-Fock theory for Bose-Einstein
  condensates and the inclusion of correlation effects}},\ \href {\doibase
  10.1103/PhysRevA.55.1147} {\bibfield  {journal} {\bibinfo  {journal} {Phys.
  Rev. A}\ }\textbf {\bibinfo {volume} {55}}, \bibinfo {pages} {1147--1159}
  (\bibinfo {year} {1997})}\BibitemShut {NoStop}%
\bibitem [{\citenamefont {Walsworth}\ and\ \citenamefont
  {You}(1997)}]{walsworth97}%
  \BibitemOpen
  \bibfield  {author} {\bibinfo {author} {\bibfnamefont {R.}~\bibnamefont
  {Walsworth}}\ and\ \bibinfo {author} {\bibfnamefont {L.}~\bibnamefont
  {You}},\ }\emph {\bibinfo {title} {Selective creation of quasiparticles in
  trapped Bose condensates}},\ \href {\doibase 10.1103/PhysRevA.56.555}
  {\bibfield  {journal} {\bibinfo  {journal} {Phys. Rev. A}\ }\textbf {\bibinfo
  {volume} {56}}, \bibinfo {pages} {555--559} (\bibinfo {year}
  {1997})}\BibitemShut {NoStop}%
\bibitem [{\citenamefont {Takei}\ \emph {et~al.}(2016)\citenamefont {Takei},
  \citenamefont {Sommer}, \citenamefont {Genes}, \citenamefont {Pupillo},
  \citenamefont {Goto}, \citenamefont {Koyasu}, \citenamefont {Chiba},
  \citenamefont {Weidem{\"u}ller},\ and\ \citenamefont
  {Ohmori}}]{ultrafastrydberg}%
  \BibitemOpen
  \bibfield  {author} {\bibinfo {author} {\bibfnamefont {N.}~\bibnamefont
  {Takei}}, \bibinfo {author} {\bibfnamefont {C.}~\bibnamefont {Sommer}},
  \bibinfo {author} {\bibfnamefont {C.}~\bibnamefont {Genes}}, \bibinfo
  {author} {\bibfnamefont {G.}~\bibnamefont {Pupillo}}, \bibinfo {author}
  {\bibfnamefont {H.}~\bibnamefont {Goto}}, \bibinfo {author} {\bibfnamefont
  {K.}~\bibnamefont {Koyasu}}, \bibinfo {author} {\bibfnamefont
  {H.}~\bibnamefont {Chiba}}, \bibinfo {author} {\bibfnamefont
  {M.}~\bibnamefont {Weidem{\"u}ller}}, \ and\ \bibinfo {author} {\bibfnamefont
  {K.}~\bibnamefont {Ohmori}},\ }\emph {\bibinfo {title} {Direct observation of
  ultrafast many-body electron dynamics in an ultracold Rydberg gas}},\ \href
  {http://dx.doi.org/10.1038/ncomms13449} {\bibfield  {journal} {\bibinfo
  {journal} {Nature Communications}\ }\textbf {\bibinfo {volume} {7}}, \bibinfo
  {pages} {13449 EP --} (\bibinfo {year} {2016})}\BibitemShut {NoStop}%
\bibitem [{\citenamefont {Stellmer}\ \emph {et~al.}(2009)\citenamefont
  {Stellmer}, \citenamefont {Tey}, \citenamefont {Huang}, \citenamefont
  {Grimm},\ and\ \citenamefont {Schreck}}]{strontium84BEC}%
  \BibitemOpen
  \bibfield  {author} {\bibinfo {author} {\bibfnamefont {S.}~\bibnamefont
  {Stellmer}}, \bibinfo {author} {\bibfnamefont {M.~K.}\ \bibnamefont {Tey}},
  \bibinfo {author} {\bibfnamefont {B.}~\bibnamefont {Huang}}, \bibinfo
  {author} {\bibfnamefont {R.}~\bibnamefont {Grimm}}, \ and\ \bibinfo {author}
  {\bibfnamefont {F.}~\bibnamefont {Schreck}},\ }\emph {\bibinfo {title}
  {Bose-{E}instein Condensation of Strontium}},\ \href {\doibase
  10.1103/PhysRevLett.103.200401} {\bibfield  {journal} {\bibinfo  {journal}
  {Phys. Rev. Lett.}\ }\textbf {\bibinfo {volume} {103}}, \bibinfo {pages}
  {200401} (\bibinfo {year} {2009})}\BibitemShut {NoStop}%
\bibitem [{\citenamefont {Yarkony}(2001)}]{CI-2001}%
  \BibitemOpen
  \bibfield  {author} {\bibinfo {author} {\bibfnamefont {D.~R.}\ \bibnamefont
  {Yarkony}},\ }\emph {\bibinfo {title} {Conical Intersections: The New
  Conventional Wisdom}},\ \href {http://dx.doi.org/10.1021/jp003731u}
  {\bibfield  {journal} {\bibinfo  {journal} {J. Phys. Chem. A}\ }\textbf
  {\bibinfo {volume} {105}}, \bibinfo {pages} {6277--6293} (\bibinfo {year}
  {2001})}\BibitemShut {NoStop}%
\bibitem [{\citenamefont {Milo{\v{s}}evi{\'{c}}}\ \emph
  {et~al.}(2006)\citenamefont {Milo{\v{s}}evi{\'{c}}}, \citenamefont {Paulus},
  \citenamefont {Bauer},\ and\ \citenamefont
  {Becker}}]{fewcycleionization-zerorange}%
  \BibitemOpen
  \bibfield  {author} {\bibinfo {author} {\bibfnamefont {D.~B.}\ \bibnamefont
  {Milo{\v{s}}evi{\'{c}}}}, \bibinfo {author} {\bibfnamefont {G.~G.}\
  \bibnamefont {Paulus}}, \bibinfo {author} {\bibfnamefont {D.}~\bibnamefont
  {Bauer}}, \ and\ \bibinfo {author} {\bibfnamefont {W.}~\bibnamefont
  {Becker}},\ }\emph {\bibinfo {title} {Above-threshold ionization by few-cycle
  pulses}},\ \href {\doibase 10.1088/0953-4075/39/14/R01} {\bibfield  {journal}
  {\bibinfo  {journal} {J. Phys. B: At. Mol. Opt. Phys.}\ }\textbf {\bibinfo
  {volume} {39}}, \bibinfo {pages} {R203} (\bibinfo {year} {2006})}\BibitemShut
  {NoStop}%
\bibitem [{\citenamefont {Hentschel}\ \emph {et~al.}(2001)\citenamefont
  {Hentschel}, \citenamefont {Kienberger}, \citenamefont {Spielmann},
  \citenamefont {Reider}, \citenamefont {Milosevic}, \citenamefont {Brabec},
  \citenamefont {Corkum}, \citenamefont {Heinzmann}, \citenamefont {Drescher},\
  and\ \citenamefont {Krausz}}]{Hentschel2001}%
  \BibitemOpen
  \bibfield  {author} {\bibinfo {author} {\bibfnamefont {M.}~\bibnamefont
  {Hentschel}}, \bibinfo {author} {\bibfnamefont {R.}~\bibnamefont
  {Kienberger}}, \bibinfo {author} {\bibfnamefont {C.}~\bibnamefont
  {Spielmann}}, \bibinfo {author} {\bibfnamefont {G.~A.}\ \bibnamefont
  {Reider}}, \bibinfo {author} {\bibfnamefont {N.}~\bibnamefont {Milosevic}},
  \bibinfo {author} {\bibfnamefont {T.}~\bibnamefont {Brabec}}, \bibinfo
  {author} {\bibfnamefont {P.}~\bibnamefont {Corkum}}, \bibinfo {author}
  {\bibfnamefont {U.}~\bibnamefont {Heinzmann}}, \bibinfo {author}
  {\bibfnamefont {M.}~\bibnamefont {Drescher}}, \ and\ \bibinfo {author}
  {\bibfnamefont {F.}~\bibnamefont {Krausz}},\ }\emph {\bibinfo {title}
  {{Attosecond metrology}}},\ \href {\doibase 10.1038/35107000} {\bibfield
  {journal} {\bibinfo  {journal} {Nature (London)}\ }\textbf {\bibinfo {volume}
  {414}}, \bibinfo {pages} {509--513} (\bibinfo {year} {2001})}\BibitemShut
  {NoStop}%
\bibitem [{\citenamefont {Itatani}\ \emph {et~al.}(2002)\citenamefont
  {Itatani}, \citenamefont {Qu\'er\'e}, \citenamefont {Yudin}, \citenamefont
  {Ivanov}, \citenamefont {Krausz},\ and\ \citenamefont
  {Corkum}}]{Itatani2002}%
  \BibitemOpen
  \bibfield  {author} {\bibinfo {author} {\bibfnamefont {J.}~\bibnamefont
  {Itatani}}, \bibinfo {author} {\bibfnamefont {F.}~\bibnamefont {Qu\'er\'e}},
  \bibinfo {author} {\bibfnamefont {G.~L.}\ \bibnamefont {Yudin}}, \bibinfo
  {author} {\bibfnamefont {M.~Y.}\ \bibnamefont {Ivanov}}, \bibinfo {author}
  {\bibfnamefont {F.}~\bibnamefont {Krausz}}, \ and\ \bibinfo {author}
  {\bibfnamefont {P.~B.}\ \bibnamefont {Corkum}},\ }\emph {\bibinfo {title}
  {Attosecond Streak Camera}},\ \href {\doibase 10.1103/PhysRevLett.88.173903}
  {\bibfield  {journal} {\bibinfo  {journal} {Phys. Rev. Lett.}\ }\textbf
  {\bibinfo {volume} {88}}, \bibinfo {pages} {173903} (\bibinfo {year}
  {2002})}\BibitemShut {NoStop}%
\bibitem [{\citenamefont {Krausz}\ and\ \citenamefont
  {Ivanov}(2009{\natexlab{b}})}]{Krausz2009}%
  \BibitemOpen
  \bibfield  {author} {\bibinfo {author} {\bibfnamefont {F.}~\bibnamefont
  {Krausz}}\ and\ \bibinfo {author} {\bibfnamefont {M.}~\bibnamefont
  {Ivanov}},\ }\emph {\bibinfo {title} {Attosecond physics}},\ \href {\doibase
  10.1103/RevModPhys.81.163} {\bibfield  {journal} {\bibinfo  {journal} {Rev.
  Mod. Phys.}\ }\textbf {\bibinfo {volume} {81}}, \bibinfo {pages} {163--234}
  (\bibinfo {year} {2009}{\natexlab{b}})}\BibitemShut {NoStop}%
\bibitem [{\citenamefont {Chin}\ and\ \citenamefont {Isenor}(1970)}]{Chin1970}%
  \BibitemOpen
  \bibfield  {author} {\bibinfo {author} {\bibfnamefont {S.~L.}\ \bibnamefont
  {Chin}}\ and\ \bibinfo {author} {\bibfnamefont {N.~R.}\ \bibnamefont
  {Isenor}},\ }\emph {\bibinfo {title} {Multiphoton ionization in atomic gases
  with depletion of neutral atoms}},\ \href {\doibase 10.1139/p70-183}
  {\bibfield  {journal} {\bibinfo  {journal} {Canadian Journal of Physics}\
  }\textbf {\bibinfo {volume} {48}}, \bibinfo {pages} {1445--1447} (\bibinfo
  {year} {1970})}\BibitemShut {NoStop}%
\bibitem [{\citenamefont {Mainfray}\ and\ \citenamefont
  {Manus}(1984)}]{Mainfray1984}%
  \BibitemOpen
  \bibfield  {author} {\bibinfo {author} {\bibfnamefont {G.}~\bibnamefont
  {Mainfray}}\ and\ \bibinfo {author} {\bibfnamefont {C.}~\bibnamefont
  {Manus}},\ }in\ \href@noop {} {\emph {\bibinfo {booktitle} {Multiphoton
  ionization of Atoms}}},\ \bibinfo {editor} {edited by\ \bibinfo {editor}
  {\bibfnamefont {S.~L.}\ \bibnamefont {Chin}}\ and\ \bibinfo {editor}
  {\bibfnamefont {P.}~\bibnamefont {Lambropoulos}}}\ (\bibinfo  {publisher}
  {Academic Press, Toronto},\ \bibinfo {year} {1984})\ pp.\ \bibinfo {pages}
  {7--34}\BibitemShut {NoStop}%
\bibitem [{\citenamefont {l'Huillier}\ \emph {et~al.}(1983)\citenamefont
  {l'Huillier}, \citenamefont {Lompre}, \citenamefont {Mainfray},\ and\
  \citenamefont {Manus}}]{LHuillier1983}%
  \BibitemOpen
  \bibfield  {author} {\bibinfo {author} {\bibfnamefont {A.}~\bibnamefont
  {l'Huillier}}, \bibinfo {author} {\bibfnamefont {L.~A.}\ \bibnamefont
  {Lompre}}, \bibinfo {author} {\bibfnamefont {G.}~\bibnamefont {Mainfray}}, \
  and\ \bibinfo {author} {\bibfnamefont {C.}~\bibnamefont {Manus}},\ }\emph
  {\bibinfo {title} {Multiply charged ions induced by multiphoton absorption in
  rare gases at 0.53 \ensuremath{\mu}m}},\ \href {\doibase
  10.1103/PhysRevA.27.2503} {\bibfield  {journal} {\bibinfo  {journal} {Phys.
  Rev. A}\ }\textbf {\bibinfo {volume} {27}}, \bibinfo {pages} {2503--2512}
  (\bibinfo {year} {1983})}\BibitemShut {NoStop}%
\bibitem [{\citenamefont {Baltu{\v{s}}ka}\ \emph {et~al.}(2003)\citenamefont
  {Baltu{\v{s}}ka}, \citenamefont {Udem}, \citenamefont {Uiberacker},
  \citenamefont {Hentschel}, \citenamefont {Goulielmakis}, \citenamefont
  {Gohle}, \citenamefont {Holzwarth}, \citenamefont {Yakovlev}, \citenamefont
  {Scrinzi}, \citenamefont {H{\"{a}}nsch},\ and\ \citenamefont
  {Krausz}}]{Baltuska2003}%
  \BibitemOpen
  \bibfield  {author} {\bibinfo {author} {\bibfnamefont {A.}~\bibnamefont
  {Baltu{\v{s}}ka}}, \bibinfo {author} {\bibfnamefont {T.}~\bibnamefont
  {Udem}}, \bibinfo {author} {\bibfnamefont {M.}~\bibnamefont {Uiberacker}},
  \bibinfo {author} {\bibfnamefont {M.}~\bibnamefont {Hentschel}}, \bibinfo
  {author} {\bibfnamefont {E.}~\bibnamefont {Goulielmakis}}, \bibinfo {author}
  {\bibfnamefont {C.}~\bibnamefont {Gohle}}, \bibinfo {author} {\bibfnamefont
  {R.}~\bibnamefont {Holzwarth}}, \bibinfo {author} {\bibfnamefont {V.~S.}\
  \bibnamefont {Yakovlev}}, \bibinfo {author} {\bibfnamefont {A.}~\bibnamefont
  {Scrinzi}}, \bibinfo {author} {\bibfnamefont {T.~W.}\ \bibnamefont
  {H{\"{a}}nsch}}, \ and\ \bibinfo {author} {\bibfnamefont {F.}~\bibnamefont
  {Krausz}},\ }\emph {\bibinfo {title} {{Attosecond control of electronic
  processes by intense light fields}}},\ \href {\doibase 10.1038/nature01414}
  {\bibfield  {journal} {\bibinfo  {journal} {Nature (London)}\ }\textbf
  {\bibinfo {volume} {421}}, \bibinfo {pages} {611--615} (\bibinfo {year}
  {2003})}\BibitemShut {NoStop}%
\bibitem [{\citenamefont {Peng}\ and\ \citenamefont
  {Starace}(2007)}]{Peng2007}%
  \BibitemOpen
  \bibfield  {author} {\bibinfo {author} {\bibfnamefont {L.-Y.}\ \bibnamefont
  {Peng}}\ and\ \bibinfo {author} {\bibfnamefont {A.~F.}\ \bibnamefont
  {Starace}},\ }\emph {\bibinfo {title} {{Attosecond pulse carrier-envelope
  phase effects on ionized electron momentum and energy distributions}}},\
  \href {\doibase 10.1103/PhysRevA.76.043401} {\bibfield  {journal} {\bibinfo
  {journal} {Phys. Rev. A}\ }\textbf {\bibinfo {volume} {76}}, \bibinfo {pages}
  {043401} (\bibinfo {year} {2007})}\BibitemShut {NoStop}%
\bibitem [{\citenamefont {Kling}\ \emph {et~al.}(2008)\citenamefont {Kling},
  \citenamefont {Rauschenberger}, \citenamefont {Verhoef}, \citenamefont
  {Hasov{\'{i}}}, \citenamefont {Uphues}, \citenamefont {{Mil Sev{\'{i}} C}},
  \citenamefont {Muller},\ and\ \citenamefont {Vrakking}}]{Kling2008}%
  \BibitemOpen
  \bibfield  {author} {\bibinfo {author} {\bibfnamefont {M.~F.}\ \bibnamefont
  {Kling}}, \bibinfo {author} {\bibfnamefont {J.}~\bibnamefont
  {Rauschenberger}}, \bibinfo {author} {\bibfnamefont {A.~J.}\ \bibnamefont
  {Verhoef}}, \bibinfo {author} {\bibfnamefont {E.}~\bibnamefont
  {Hasov{\'{i}}}}, \bibinfo {author} {\bibfnamefont {T.}~\bibnamefont
  {Uphues}}, \bibinfo {author} {\bibfnamefont {D.~B.}\ \bibnamefont {{Mil
  Sev{\'{i}} C}}}, \bibinfo {author} {\bibfnamefont {H.~G.}\ \bibnamefont
  {Muller}}, \ and\ \bibinfo {author} {\bibfnamefont {M.~J.~J.}\ \bibnamefont
  {Vrakking}},\ }\emph {\bibinfo {title} {{Imaging of carrier-envelope phase
  effects in above-threshold ionization with intense few-cycle laser
  fields}}},\ \href {\doibase 10.1088/1367-2630/10/2/025024} {\bibfield
  {journal} {\bibinfo  {journal} {New J. Phys.}\ }\textbf {\bibinfo {volume}
  {10}}, \bibinfo {pages} {25024--25024} (\bibinfo {year} {2008})}\BibitemShut
  {NoStop}%
\bibitem [{\citenamefont {Kr{\"{u}}ger}\ \emph {et~al.}(2011)\citenamefont
  {Kr{\"{u}}ger}, \citenamefont {Schenk},\ and\ \citenamefont
  {Hommelhoff}}]{Kruger2011}%
  \BibitemOpen
  \bibfield  {author} {\bibinfo {author} {\bibfnamefont {M.}~\bibnamefont
  {Kr{\"{u}}ger}}, \bibinfo {author} {\bibfnamefont {M.}~\bibnamefont
  {Schenk}}, \ and\ \bibinfo {author} {\bibfnamefont {P.}~\bibnamefont
  {Hommelhoff}},\ }\emph {\bibinfo {title} {{Attosecond control of electrons
  emitted from a nanoscale metal tip}}},\ \href {\doibase 10.1038/nature10196}
  {\bibfield  {journal} {\bibinfo  {journal} {Nature (London)}\ }\textbf
  {\bibinfo {volume} {475}}, \bibinfo {pages} {78--81} (\bibinfo {year}
  {2011})}\BibitemShut {NoStop}%
\bibitem [{\citenamefont {Kling}\ \emph {et~al.}(2006)\citenamefont {Kling},
  \citenamefont {Siedschlag}, \citenamefont {Verhoef}, \citenamefont {Khan},
  \citenamefont {Schultze}, \citenamefont {Uphues}, \citenamefont {Ni},
  \citenamefont {Uiberacker}, \citenamefont {Drescher}, \citenamefont
  {Krausz},\ and\ \citenamefont {Vrakking}}]{Kling2006}%
  \BibitemOpen
  \bibfield  {author} {\bibinfo {author} {\bibfnamefont {M.~F.}\ \bibnamefont
  {Kling}}, \bibinfo {author} {\bibfnamefont {C.}~\bibnamefont {Siedschlag}},
  \bibinfo {author} {\bibfnamefont {A.~J.}\ \bibnamefont {Verhoef}}, \bibinfo
  {author} {\bibfnamefont {J.~I.}\ \bibnamefont {Khan}}, \bibinfo {author}
  {\bibfnamefont {M.}~\bibnamefont {Schultze}}, \bibinfo {author}
  {\bibfnamefont {T.}~\bibnamefont {Uphues}}, \bibinfo {author} {\bibfnamefont
  {Y.}~\bibnamefont {Ni}}, \bibinfo {author} {\bibfnamefont {M.}~\bibnamefont
  {Uiberacker}}, \bibinfo {author} {\bibfnamefont {M.}~\bibnamefont
  {Drescher}}, \bibinfo {author} {\bibfnamefont {F.}~\bibnamefont {Krausz}}, \
  and\ \bibinfo {author} {\bibfnamefont {M.~J.~J.}\ \bibnamefont {Vrakking}},\
  }\emph {\bibinfo {title} {{Control of electron localization in molecular
  dissociation.}}},\ \href {\doibase 10.1126/science.1126259} {\bibfield
  {journal} {\bibinfo  {journal} {Science}\ }\textbf {\bibinfo {volume} {312}},
  \bibinfo {pages} {246--8} (\bibinfo {year} {2006})}\BibitemShut {NoStop}%
\bibitem [{\citenamefont {Rathje}\ \emph {et~al.}(2013)\citenamefont {Rathje},
  \citenamefont {Sayler}, \citenamefont {Zeng}, \citenamefont {Wustelt},
  \citenamefont {Figger}, \citenamefont {Esry},\ and\ \citenamefont
  {Paulus}}]{Rathje2013}%
  \BibitemOpen
  \bibfield  {author} {\bibinfo {author} {\bibfnamefont {T.}~\bibnamefont
  {Rathje}}, \bibinfo {author} {\bibfnamefont {A.~M.}\ \bibnamefont {Sayler}},
  \bibinfo {author} {\bibfnamefont {S.}~\bibnamefont {Zeng}}, \bibinfo {author}
  {\bibfnamefont {P.}~\bibnamefont {Wustelt}}, \bibinfo {author} {\bibfnamefont
  {H.}~\bibnamefont {Figger}}, \bibinfo {author} {\bibfnamefont {B.~D.}\
  \bibnamefont {Esry}}, \ and\ \bibinfo {author} {\bibfnamefont {G.~G.}\
  \bibnamefont {Paulus}},\ }\emph {\bibinfo {title} {{Coherent Control at Its
  Most Fundamental: Carrier-Envelope-Phase-Dependent Electron Localization in
  Photodissociation of a H þ 2 Molecular Ion Beam Target}}},\ \href {\doibase
  10.1103/PhysRevLett.111.093002} {\bibfield  {journal} {\bibinfo  {journal}
  {Phys. Rev. Lett.}\ }\textbf {\bibinfo {volume} {111}}, \bibinfo {pages}
  {093002} (\bibinfo {year} {2013})}\BibitemShut {NoStop}%
\bibitem [{\citenamefont {Ngoko~Djiokap}\ \emph {et~al.}(2015)\citenamefont
  {Ngoko~Djiokap}, \citenamefont {Hu}, \citenamefont {Madsen}, \citenamefont
  {Manakov}, \citenamefont {Meremianin},\ and\ \citenamefont
  {Starace}}]{starace-polarization}%
  \BibitemOpen
  \bibfield  {author} {\bibinfo {author} {\bibfnamefont {J.~M.}\ \bibnamefont
  {Ngoko~Djiokap}}, \bibinfo {author} {\bibfnamefont {S.~X.}\ \bibnamefont
  {Hu}}, \bibinfo {author} {\bibfnamefont {L.~B.}\ \bibnamefont {Madsen}},
  \bibinfo {author} {\bibfnamefont {N.~L.}\ \bibnamefont {Manakov}}, \bibinfo
  {author} {\bibfnamefont {A.~V.}\ \bibnamefont {Meremianin}}, \ and\ \bibinfo
  {author} {\bibfnamefont {A.~F.}\ \bibnamefont {Starace}},\ }\emph {\bibinfo
  {title} {Electron Vortices in Photoionization by Circularly Polarized
  Attosecond Pulses}},\ \href {\doibase 10.1103/PhysRevLett.115.113004}
  {\bibfield  {journal} {\bibinfo  {journal} {Phys. Rev. Lett.}\ }\textbf
  {\bibinfo {volume} {115}}, \bibinfo {pages} {113004} (\bibinfo {year}
  {2015})}\BibitemShut {NoStop}%
\bibitem [{\citenamefont {Pont}\ and\ \citenamefont
  {Gavrila}(1990)}]{stronfieldstabilization}%
  \BibitemOpen
  \bibfield  {author} {\bibinfo {author} {\bibfnamefont {M.}~\bibnamefont
  {Pont}}\ and\ \bibinfo {author} {\bibfnamefont {M.}~\bibnamefont {Gavrila}},\
  }\emph {\bibinfo {title} {Stabilization of atomic hydrogen in superintense,
  high-frequency laser fields of circular polarization}},\ \href {\doibase
  10.1103/PhysRevLett.65.2362} {\bibfield  {journal} {\bibinfo  {journal}
  {Phys. Rev. Lett.}\ }\textbf {\bibinfo {volume} {65}}, \bibinfo {pages}
  {2362--2365} (\bibinfo {year} {1990})}\BibitemShut {NoStop}%
\bibitem [{\citenamefont {Shafir}\ \emph {et~al.}(2012)\citenamefont {Shafir},
  \citenamefont {Soifer}, \citenamefont {Bruner}, \citenamefont {Dagan},
  \citenamefont {Mairesse}, \citenamefont {Patchkovskii}, \citenamefont
  {Ivanov}, \citenamefont {Smirnova},\ and\ \citenamefont
  {Dudovich}}]{dudovich-tunnelingtime}%
  \BibitemOpen
  \bibfield  {author} {\bibinfo {author} {\bibfnamefont {D.}~\bibnamefont
  {Shafir}}, \bibinfo {author} {\bibfnamefont {H.}~\bibnamefont {Soifer}},
  \bibinfo {author} {\bibfnamefont {B.~D.}\ \bibnamefont {Bruner}}, \bibinfo
  {author} {\bibfnamefont {M.}~\bibnamefont {Dagan}}, \bibinfo {author}
  {\bibfnamefont {Y.}~\bibnamefont {Mairesse}}, \bibinfo {author}
  {\bibfnamefont {S.}~\bibnamefont {Patchkovskii}}, \bibinfo {author}
  {\bibfnamefont {M.~Y.}\ \bibnamefont {Ivanov}}, \bibinfo {author}
  {\bibfnamefont {O.}~\bibnamefont {Smirnova}}, \ and\ \bibinfo {author}
  {\bibfnamefont {N.}~\bibnamefont {Dudovich}},\ }\emph {\bibinfo {title}
  {Resolving the time when an electron exits a tunnelling barrier}},\ \href
  {http://dx.doi.org/10.1038/nature11025} {\bibfield  {journal} {\bibinfo
  {journal} {Nature (London)}\ }\textbf {\bibinfo {volume} {485}}, \bibinfo
  {pages} {343--346} (\bibinfo {year} {2012})}\BibitemShut {NoStop}%
\bibitem [{\citenamefont {Pfeiffer}\ \emph {et~al.}(2012)\citenamefont
  {Pfeiffer}, \citenamefont {Cirelli}, \citenamefont {Smolarski}, \citenamefont
  {Dimitrovski}, \citenamefont {Abu-samha}, \citenamefont {Madsen},\ and\
  \citenamefont {Keller}}]{attoclock-tunnelingtiming}%
  \BibitemOpen
  \bibfield  {author} {\bibinfo {author} {\bibfnamefont {A.~N.}\ \bibnamefont
  {Pfeiffer}}, \bibinfo {author} {\bibfnamefont {C.}~\bibnamefont {Cirelli}},
  \bibinfo {author} {\bibfnamefont {M.}~\bibnamefont {Smolarski}}, \bibinfo
  {author} {\bibfnamefont {D.}~\bibnamefont {Dimitrovski}}, \bibinfo {author}
  {\bibfnamefont {M.}~\bibnamefont {Abu-samha}}, \bibinfo {author}
  {\bibfnamefont {L.~B.}\ \bibnamefont {Madsen}}, \ and\ \bibinfo {author}
  {\bibfnamefont {U.}~\bibnamefont {Keller}},\ }\emph {\bibinfo {title}
  {Attoclock reveals natural coordinates of the laser-induced tunnelling
  current flow in atoms}},\ \href {http://dx.doi.org/10.1038/nphys2125}
  {\bibfield  {journal} {\bibinfo  {journal} {Nat. Phys.}\ }\textbf {\bibinfo
  {volume} {8}}, \bibinfo {pages} {76--80} (\bibinfo {year}
  {2012})}\BibitemShut {NoStop}%
\bibitem [{\citenamefont {Uiberacker}\ \emph {et~al.}(2007)\citenamefont
  {Uiberacker}, \citenamefont {Uphues}, \citenamefont {Schultze}, \citenamefont
  {Verhoef}, \citenamefont {Yakovlev}, \citenamefont {Kling}, \citenamefont
  {Rauschenberger}, \citenamefont {Kabachnik}, \citenamefont {Schroder},
  \citenamefont {Lezius}, \citenamefont {Kompa}, \citenamefont {Muller},
  \citenamefont {Vrakking}, \citenamefont {Hendel}, \citenamefont {Kleineberg},
  \citenamefont {Heinzmann}, \citenamefont {Drescher},\ and\ \citenamefont
  {Krausz}}]{krausz-tunnelingtime}%
  \BibitemOpen
  \bibfield  {author} {\bibinfo {author} {\bibfnamefont {M.}~\bibnamefont
  {Uiberacker}}, \bibinfo {author} {\bibfnamefont {T.}~\bibnamefont {Uphues}},
  \bibinfo {author} {\bibfnamefont {M.}~\bibnamefont {Schultze}}, \bibinfo
  {author} {\bibfnamefont {A.~J.}\ \bibnamefont {Verhoef}}, \bibinfo {author}
  {\bibfnamefont {V.}~\bibnamefont {Yakovlev}}, \bibinfo {author}
  {\bibfnamefont {M.~F.}\ \bibnamefont {Kling}}, \bibinfo {author}
  {\bibfnamefont {J.}~\bibnamefont {Rauschenberger}}, \bibinfo {author}
  {\bibfnamefont {N.~M.}\ \bibnamefont {Kabachnik}}, \bibinfo {author}
  {\bibfnamefont {H.}~\bibnamefont {Schroder}}, \bibinfo {author}
  {\bibfnamefont {M.}~\bibnamefont {Lezius}}, \bibinfo {author} {\bibfnamefont
  {K.~L.}\ \bibnamefont {Kompa}}, \bibinfo {author} {\bibfnamefont {H.~G.}\
  \bibnamefont {Muller}}, \bibinfo {author} {\bibfnamefont {M.~J.~J.}\
  \bibnamefont {Vrakking}}, \bibinfo {author} {\bibfnamefont {S.}~\bibnamefont
  {Hendel}}, \bibinfo {author} {\bibfnamefont {U.}~\bibnamefont {Kleineberg}},
  \bibinfo {author} {\bibfnamefont {U.}~\bibnamefont {Heinzmann}}, \bibinfo
  {author} {\bibfnamefont {M.}~\bibnamefont {Drescher}}, \ and\ \bibinfo
  {author} {\bibfnamefont {F.}~\bibnamefont {Krausz}},\ }\emph {\bibinfo
  {title} {Attosecond real-time observation of electron tunnelling in atoms}},\
  \href {http://dx.doi.org/10.1038/nature05648} {\bibfield  {journal} {\bibinfo
   {journal} {Nature (London)}\ }\textbf {\bibinfo {volume} {446}}, \bibinfo
  {pages} {627--632} (\bibinfo {year} {2007})}\BibitemShut {NoStop}%
\bibitem [{\citenamefont {Isinger}\ \emph {et~al.}(2017)\citenamefont
  {Isinger}, \citenamefont {Squibb}, \citenamefont {Busto}, \citenamefont
  {Zhong}, \citenamefont {Harth}, \citenamefont {Kroon}, \citenamefont {Nandi},
  \citenamefont {Arnold}, \citenamefont {Miranda}, \citenamefont
  {Dahlstr{\"o}m}, \citenamefont {Lindroth}, \citenamefont {Feifel},
  \citenamefont {Gisselbrecht},\ and\ \citenamefont
  {L{\textquoteright}Huillier}}]{lindrothhuillierphotoionization}%
  \BibitemOpen
  \bibfield  {author} {\bibinfo {author} {\bibfnamefont {M.}~\bibnamefont
  {Isinger}}, \bibinfo {author} {\bibfnamefont {R.~J.}\ \bibnamefont {Squibb}},
  \bibinfo {author} {\bibfnamefont {D.}~\bibnamefont {Busto}}, \bibinfo
  {author} {\bibfnamefont {S.}~\bibnamefont {Zhong}}, \bibinfo {author}
  {\bibfnamefont {A.}~\bibnamefont {Harth}}, \bibinfo {author} {\bibfnamefont
  {D.}~\bibnamefont {Kroon}}, \bibinfo {author} {\bibfnamefont
  {S.}~\bibnamefont {Nandi}}, \bibinfo {author} {\bibfnamefont {C.~L.}\
  \bibnamefont {Arnold}}, \bibinfo {author} {\bibfnamefont {M.}~\bibnamefont
  {Miranda}}, \bibinfo {author} {\bibfnamefont {J.~M.}\ \bibnamefont
  {Dahlstr{\"o}m}}, \bibinfo {author} {\bibfnamefont {E.}~\bibnamefont
  {Lindroth}}, \bibinfo {author} {\bibfnamefont {R.}~\bibnamefont {Feifel}},
  \bibinfo {author} {\bibfnamefont {M.}~\bibnamefont {Gisselbrecht}}, \ and\
  \bibinfo {author} {\bibfnamefont {A.}~\bibnamefont
  {L{\textquoteright}Huillier}},\ }\emph {\bibinfo {title} {Photoionization in
  the time and frequency domain}},\ \href {\doibase 10.1126/science.aao7043}
  {\bibfield  {journal} {\bibinfo  {journal} {Science}\ }\textbf {\bibinfo
  {volume} {358}}, \bibinfo {pages} {893--896} (\bibinfo {year}
  {2017})}\BibitemShut {NoStop}%
\bibitem [{\citenamefont {Serwane}\ \emph {et~al.}(2011)\citenamefont
  {Serwane}, \citenamefont {Z{\"u}rn}, \citenamefont {Lompe}, \citenamefont
  {Ottenstein}, \citenamefont {Wenz},\ and\ \citenamefont
  {Jochim}}]{jochim-fewfermion}%
  \BibitemOpen
  \bibfield  {author} {\bibinfo {author} {\bibfnamefont {F.}~\bibnamefont
  {Serwane}}, \bibinfo {author} {\bibfnamefont {G.}~\bibnamefont {Z{\"u}rn}},
  \bibinfo {author} {\bibfnamefont {T.}~\bibnamefont {Lompe}}, \bibinfo
  {author} {\bibfnamefont {T.~B.}\ \bibnamefont {Ottenstein}}, \bibinfo
  {author} {\bibfnamefont {A.~N.}\ \bibnamefont {Wenz}}, \ and\ \bibinfo
  {author} {\bibfnamefont {S.}~\bibnamefont {Jochim}},\ }\emph {\bibinfo
  {title} {Deterministic Preparation of a Tunable Few-Fermion System}},\ \href
  {http://dx.doi.org/10.1126/science.1201351} {\bibfield  {journal} {\bibinfo
  {journal} {Science}\ }\textbf {\bibinfo {volume} {332}}, \bibinfo {pages}
  {336--338} (\bibinfo {year} {2011})}\BibitemShut {NoStop}%
\bibitem [{\citenamefont {Zinner}\ and\ \citenamefont
  {Jensen}(2013)}]{zinner-nuclearQS}%
  \BibitemOpen
  \bibfield  {author} {\bibinfo {author} {\bibfnamefont {N.~T.}\ \bibnamefont
  {Zinner}}\ and\ \bibinfo {author} {\bibfnamefont {A.~S.}\ \bibnamefont
  {Jensen}},\ }\emph {\bibinfo {title} {Comparing and contrasting nuclei and
  cold atomic gases}},\ \href {\doibase 10.1088/0954-3899/40/5/053101}
  {\bibfield  {journal} {\bibinfo  {journal} {Journal of Physics G: Nuclear and
  Particle Physics}\ }\textbf {\bibinfo {volume} {40}}, \bibinfo {pages}
  {053101} (\bibinfo {year} {2013})}\BibitemShut {NoStop}%
\bibitem [{\citenamefont {Keshet}\ and\ \citenamefont
  {Ketterle}(2013)}]{Cicero}%
  \BibitemOpen
  \bibfield  {author} {\bibinfo {author} {\bibfnamefont {A.}~\bibnamefont
  {Keshet}}\ and\ \bibinfo {author} {\bibfnamefont {W.}~\bibnamefont
  {Ketterle}},\ }\emph {\bibinfo {title} {A distributed, graphical user
  interface based, computer control system for atomic physics experiments}},\
  \href {\doibase 10.1063/1.4773536} {\bibfield  {journal} {\bibinfo  {journal}
  {Rev. Sci. Instrum.}\ }\textbf {\bibinfo {volume} {84}}, \bibinfo {pages}
  {015105} (\bibinfo {year} {2013})}\BibitemShut {NoStop}%
\bibitem [{\citenamefont {Senaratne}\ \emph {et~al.}(2015)\citenamefont
  {Senaratne}, \citenamefont {Rajagopal}, \citenamefont {Geiger}, \citenamefont
  {Fujiwara}, \citenamefont {Lebedev},\ and\ \citenamefont {Weld}}]{nozzleRSI}%
  \BibitemOpen
  \bibfield  {author} {\bibinfo {author} {\bibfnamefont {R.}~\bibnamefont
  {Senaratne}}, \bibinfo {author} {\bibfnamefont {S.~V.}\ \bibnamefont
  {Rajagopal}}, \bibinfo {author} {\bibfnamefont {Z.~A.}\ \bibnamefont
  {Geiger}}, \bibinfo {author} {\bibfnamefont {K.~M.}\ \bibnamefont
  {Fujiwara}}, \bibinfo {author} {\bibfnamefont {V.}~\bibnamefont {Lebedev}}, \
  and\ \bibinfo {author} {\bibfnamefont {D.~M.}\ \bibnamefont {Weld}},\ }\emph
  {\bibinfo {title} {Effusive atomic oven nozzle design using an aligned
  microcapillary array}},\ \href {\doibase http://dx.doi.org/10.1063/1.4907401}
  {\bibfield  {journal} {\bibinfo  {journal} {Rev. Sci. Instrum.}\ }\textbf
  {\bibinfo {volume} {86}},\ \bibinfo {eid} {023105} (\bibinfo {year}
  {2015})}\BibitemShut {NoStop}%
\end{thebibliography}
\end{document}